\documentclass[twocolumn]{aastex63}
\submitjournal{ApJ}

\usepackage{amsmath}

\usepackage{comment}

\graphicspath{{./}{figures/}}

\begin{document}

\correspondingauthor{H.~Ashkar, R. Konno, H. Prokoph,  F. Schüssler and S. J. Zhu}
\email{contact.hess@hess-experiment.eu}

\collaboration{0}{H.E.S.S. Collaboration}

\author{H.~Abdalla}
\affiliation{University of Namibia, Department of Physics, Private Bag 13301, Windhoek 10005, Namibia}

\author[0000-0003-1157-3915]{F.~Aharonian}
\affiliation{Dublin Institute for Advanced Studies, 31 Fitzwilliam Place, Dublin 2, Ireland}
\affiliation{Max-Planck-Institut f\"ur Kernphysik, P.O. Box 103980, D 69029 Heidelberg, Germany}
\affiliation{High Energy Astrophysics Laboratory, RAU,  123 Hovsep Emin St  Yerevan 0051, Armenia}

\author{F.~Ait~Benkhali}
\affiliation{Max-Planck-Institut f\"ur Kernphysik, P.O. Box 103980, D 69029 Heidelberg, Germany}

\author{E.O.~Ang\"uner}
\affiliation{Aix Marseille Universit\'e, CNRS/IN2P3, CPPM, Marseille, France}

\author[0000-0002-2153-1818]{H.~Ashkar}
\affiliation{IRFU, CEA, Universit\'e Paris-Saclay, F-91191 Gif-sur-Yvette, France}

\author[0000-0002-9326-6400]{M.~Backes}
\affiliation{University of Namibia, Department of Physics, Private Bag 13301, Windhoek 10005, Namibia}
\affiliation{Centre for Space Research, North-West University, Potchefstroom 2520, South Africa}

\author{V.~Baghmanyan}
\affiliation{Instytut Fizyki J\c{a}drowej PAN, ul. Radzikowskiego 152, 31-342 Krak{\'o}w, Poland}

\author[0000-0002-5085-8828]{V.~Barbosa~Martins}
\affiliation{DESY, D-15738 Zeuthen, Germany}

\author[0000-0002-5797-3386]{R.~Batzofin}
\affiliation{School of Physics, University of the Witwatersrand, 1 Jan Smuts Avenue, Braamfontein, Johannesburg, 2050 South Africa}

\author{Y.~Becherini}
\affiliation{Department of Physics and Electrical Engineering, Linnaeus University,  351 95 V\"axj\"o, Sweden}

% ...

\title{H.E.S.S. follow-up observations of Binary Black Hole Coalescence events during the second and third Gravitational Waves observing runs of Advanced LIGO and Advanced Virgo}



\begin{abstract}

We report on the observations of four well-localized binary black hole (BBH) mergers by the High Energy Stereoscopic System (H.E.S.S.) during the second and third observing runs of Advanced LIGO and Advanced Virgo, O2 and O3. H.E.S.S. can observe $\mathrm{20\,deg^2}$ of the sky at a time and follows up gravitational-wave (GW) events by ``tiling'' localization regions to maximize the covered localization probability. During O2 and O3, H.E.S.S. observed large portions of the localization regions, between 35\% and 75\%, for four BBH mergers (GW170814, GW190512\_180714, GW190728\_064510, and S200224ca). For these four GW events, we find no significant signal from a pointlike source in any of the observations, and set upper limits on the very high energy ($>$100 GeV) $\gamma$-ray emission. The $1-10$~TeV isotropic luminosity of these GW events is below $10^{45}$ erg\,s$^{-1}$ at the times of the H.E.S.S. observations, around the level of the low-luminosity GRB 190829A. Assuming no changes are made to how follow-up observations are conducted, H.E.S.S. can expect to observe over 60 GW events per year in the fourth GW observing run, O4, of which eight would be observable with minimal latency.
\end{abstract}


\section{Introduction}
\label{sec:introduction}

In 2017, the detection of both gravitational waves (GWs) and electromagnetic radiation from the merger of two neutron stars (NSs), GW170817, revolutionized multimessenger astronomy \citep{GW170817MM}. The electromagnetic radiation was observed as a short set of flashes of low-energy $\gamma$-rays, which is commonly referred to as a short gamma-ray burst (GRB). This single event solved a decades-old mystery of high-energy astrophysics by confirming that some short GRBs are produced by the merger of two compact objects, with at least one being an NS. The other population of GRBs, long GRBs, have long been associated with core-collapse supernovae and are known to arise from the collapses of the most massive stars. In either GRB progenitor scenario, the catastrophic event launches an ultrarelativistic jet. Interactions within the jet produce the highly variable prompt emission, while the jet's later interactions with the surrounding material cause it to decelerate and produce long-lived afterglow emission. This emission is observed as smoothly fading emission when the GRB is viewed near the axis of the jet. In contrast, as was the case in GRB 170817A \citep{Nynka2018_170817xrayAG, Troja2020_170817xrayAG}, if the GRB is observed off-axis, the afterglow emission is instead observed to initially rise before peaking and then fading due to geometric effects.\\


Object GRB 170817A was well observed over a wide range of photon energies. With a redshift of 0.0098 \citep{Hjorth2017_170817redshift}, corresponding to a luminosity distance of 40~Mpc, its proximity meant that its emission could be observed to much later times; the X-ray afterglow, for instance, was still detectable 1000 days after the initial event \citep{Troja2020_170817xrayAG}. 
It was also observed 
by very high energy (VHE) $\gamma$-ray detectors such as the High Energy Stereoscopic System (H.E.S.S.; \citet{HESS170817}, an array of five imaging atmospheric Cerenkov telescopes that can detect $\gamma$-rays with energies greater than tens of GeV. The upper limits on VHE emission from the late-time H.E.S.S. observations provide constraints on the magnetic field of the outflow \citep{HESS170817_deep}.

Short GRBs could also be produced by the merger of an NS and a black hole (BH; see, e.g.,~\citet{Foucart_2020_NSBHreview} for a recent review). 
In contrast, a merger of two stellar mass BHs is typically not expected to produce electromagnetic radiation due to the lack of accreting material, which is considered necessary for launching a jet. 
However, it is essential to verify this assumption with observations. Indeed, the announcement of a $\gamma$-ray transient temporally coincident with the GW150914 binary black hole (BBH) merger \citep{2016ApJ...826L...6C, 2018ApJ...853L...9C} sparked much interest and controversy and inspired explanations such as a circumbinary or remnant disk \citep{Perna2016_BBHEM, Perna2019_BBHEM, Murase2016_BBHEM, Kotera2016_BBHEM, Martin2018_BBHEM}, or charged BHs \citep{Zhang2016_BBHEM, Liebling2016_BBHEM, Fraschetti2018_BBHEM}. However, no GRB-like candidate for a BBH merger has been observed since then, although other potential sources of electromagnetic emission have been discussed \citep{Graham2020_ZTFS190521g,Bartos2017_BBHEM_AGN,Stone2017_BBHEM_AGN}.\\ 

A single interferometer has a broad antenna pattern and hence poor directional sensitivity; therefore, the localization of transient GW events with only one detector can encompass large parts of the sky. However, the subset of events that are detected by more than two GW instruments are localized to tens of square degrees or less \citep{Pankow2018_GWlocs}. H.E.S.S. has a field of view (FoV) of about 20 deg$^2$, and can observe large portions of the GW localization regions using an observational pattern known as \emph{tiling}~\citep{hessGWFollowuo_technical}, during which different parts of the sky region are observed according to some predefined sequence. During the second and third observing runs (O2 and O3) of the GW detectors Advanced LIGO and Advanced Virgo, H.E.S.S. conducted follow-up observations of six well-localized GW events, chosen so that a sky area corresponding to at least 50\% localization probability could be covered in 1 night for BBH events\footnote{An exception was made for the first event in the third observation run (O3) to test the follow-up procedures.} and at least 10\% could be covered in 1 night for mergers involving an NS. Of these six events, one was the NS\,-\,NS merger GW170817, discussed in \citet{HESS170817,HESS170817_deep}, one was an NS\,-\,BH merger, GW200105~\citep{Abbott_2021_twoNSBHs}, and the other four were BBH mergers, GW170814~\citep{GW170814}, GW190512\_180714, GW190728\_064510 ~\citep{abbott2020gwtc2} and S200224ca~\citep{S200224ca_initial}. Of these last five, the NS\,-\,BH binary was only covered by one pointing (or observation run) due to poor weather conditions, amounting to less than 1\% of the localization region; since this translates to such a small chance of having observed the correct sky region, we do not discuss the NS\,-\,BH event here and instead focus only on the BBH mergers. The same analysis methods can be used for all GW events regardless of the type of progenitor system, so that the four BBH events discussed here are a good test of H.E.S.S.'s sensitivity to VHE $\gamma$-ray emission from GW mergers. \\

The H.E.S.S. observations during O2 and O3 all began with delays of at least a few hours after the original merger events and were conducted by scanning regions of the sky to maximize the amount of the GW localization region probability that could be covered. The rate of GW events is expected to increase in the fourth GW observing run (O4), while the localization uncertainties should greatly decrease~\citep{LVK_LivingReviews}. In order to prepare for O4, it is important to assess the sensitivity obtained by the current H.E.S.S. observation strategy and determine if modifications are necessary. \\

In this paper, we present the H.E.S.S. observations of four well-localized BBH mergers. In Section~\ref{sec:observations}, we discuss the details of the observations; then, we describe the analysis of the data and report on the results in Section~\ref{sec:analysis}. We search for a pointlike source anywhere in the region covered by the observations and do not find any significant signal, so we calculate integral upper limits on the $\gamma$-ray flux. Finally, we discuss these upper limits in Section~\ref{sec:discussion}, and conclude in Section~\ref{sec:outlook}. Throughout this paper, we focus on GW190728\_064510 as an illustrative example and include the equivalent information and figures for the three other GW events in the \texttt{ADDITIONAL TABLES AND FIGURES} section. The cosmological parameters used are taken from~\citet{planck_2015} with a Hubble constant $\mathrm{H_0 = 67.8 \pm 0.9\,km s^{-1} Mpc^{-1}}$, a matter density parameter $\Omega_m = 0.308 \pm 0.012$.

\bigbreak
\section{Observations}
\label{sec:observations}

\begin{table*}[ht]
\centering
\small
\begin{tabular}{ccccccc}
  \hline
    Position & Start time (UTC) & RA J2000 (deg) & DEC J2000 (deg) &  Duration (min)  & Zenith angle (deg) \\  
        \hline
     1a & 2019-07-28 21:12 &  313.09 & 8.16  & 28 &  45 \\
     1b & 2019-07-28 21:46 &  313.09 & 8.16  & 18 &  39  \\
     2 & 2019-07-28 22:23 &  317.11 & 15.02  & 28 &  43  \\
     3a & 2019-07-28 22:59 &  314.39 & 10.81  & 14 &  35  \\
     3b & 2019-07-28 23:18 &  314.39 & 10.81  & 28 & 34  \\
     4 & 2019-07-28 23:47 &  312.98 & 5.68  & 28 &  30  \\
     5a& 2019-07-29 00:16&  318.69 & 17.19& 6 & 41   \\
     5b & 2019-07-29 00:34 &  318.69 & 17.19  & 28 &  42  \\
     6 & 2019-07-29 01:03 &  316.14 & 12.94  & 28 &  42  \\
     7 & 2019-07-29 01:32 &  312.80 & 7.03  & 28 &  44  \\
   \hline
   \hline
\end{tabular}
\caption{H.E.S.S. observations of GW190728\_064510 BBH merger GW event.}
\label{tab:HESS_OBS1}
\end{table*} 

H.E.S.S. is located in the Khomas Highland in Namibia. It consists of four 12 m telescopes with 20 deg$^2$ FoVs and one 28 m telescope with a 9 deg$^2$ FoV and is capable of detecting VHE $\gamma$ rays ranging from a few tens of GeV to 100 TeV in energy.

The H.E.S.S. Transients Alert System~\citep[in preparation]{hessTOOsystem} monitors GW alerts via the automated Gammaray burst Coordinate Network (GCN;~\citet{GCN} notices provided by the LIGO and Virgo Collaborations. These notices contain GW localization maps in HEALPix~\citep{healpy2,healpy1} format. The pixels of the map contain four layers of information: a layer containing the probability of finding the GW event in the corresponding position in the sky and three layers containing distance information. The BBH events followed by H.E.S.S. are at large distances of several hundreds to thousands of megaparsecs. Current galaxy catalogs, such as, for example, GLADE~\citep{dalya2018glade1} are not complete at these distances and cannot be used to further optimize the follow-up strategy. Therefore, the BBH events are followed by 2D grading schemes that cover the regions in the sky that have the greatest 2D probability integrated inside the H.E.S.S. FoV without taking into account distance information. By default, each region is observed with one run.

The most probable regions to host the GW event are observed first, taking into consideration observation and visibility constraints.  Imaging atmospheric Cerenkov telescopes like H.E.S.S. observe during astronomical nights with low levels of light falling into the cameras. To ensure this, the Sun and Moon altitude, Moon phase, and Moon-to-source separation need to be considered. Moreover, due to the absorption of Cerenkov shower light in the atmosphere, the energy threshold of the H.E.S.S. observations depends heavily on the respective zenith angle. In general, low zenith angle observations are favored and observations are restricted to zenith angles below 60$^\circ$. This constraint on zenith angle combined with the telescope location on Earth determines the sky visibility for H.E.S.S. at a given time.

The GW follow-ups are adapted to these H.E.S.S. visibility and observation conditions in order to observe the GW events as quickly and efficiently as possible. More details on the pointing patterns of H.E.S.S. GW follow-up observations, the description of the GW follow-up triggering framework, and the GW follow-up tools integrated within the H.E.S.S Transient Follow-up System~\citep[in preparation]{hessTOOsystem} are provided in~\citet{hessGWFollowuo_technical}. 

Event GW170814~\citep{GW170814} was the first three interferometer BBH merger detection that occurred during O2. It was also the first BBH merger whose location permitted H.E.S.S. follow-up observations. Observations were conducted over three nights (2017 August 17, 18 and 19) and are presented in~\citet{GW170814_HESS}; as this was the first tiled GW observation, this was the first opportunity to test the developed follow-up algorithms. 

During O3, H.E.S.S. performed follow-up observations of three BBH events: GW190512\_180714, GW190728\_064510 ~\citep{abbott2020gwtc2}, and S200224ca~\citep{S200224ca_initial}. In contrast to O2, the follow-up observations were restricted to a maximum delay of 24 hr and were all performed during the same night following the arrival of the GW alert. 

Event GW190512\_180714 was the first BBH to be well localized in O3 with a favorable zenith angle for H.E.S.S. The observation delay was $\sim$7.1 hr and with a total of six observation runs, the total observation duration was $\sim$2.6 hr. This GW event in particular was used to commission and streamline the H.E.S.S. response to GW alerts during O3.

Event GW190728\_064510 was initially classified as a MassGap event~\citep{S190728q_initial}. The H.E.S.S Transient System automatically scheduled follow-up observations starting with a delay of $\sim$11 hr. 
A neutrino candidate~\citep{S190728q_neutrino} event from IceCube was temporally coincident with the GW event (within 360 s) and consistent with its sky localization. When the H.E.S.S observations started, GW190728\_064510 was reclassified as a BBH merger, and the localization map was updated~\citep{S190728q_update}. A new schedule was automatically computed by the H.E.S.S. Transient System, and 10 runs were taken summing to a total of 5.65 hr of data. Follow-up observations of GW190728\_064510 covered parts of the uncertainty region of the neutrino candidate from IceCube.

Event S200224ca~\citep{S200224ca_initial} was one of the well-localised and therefore best followed-up BBH events (according to the Treasure Map\footnote{\url{http://treasuremap.space}};~\citet{TreasureMap}. H.E.S.S. took three runs covering the larger part of the localization region~\citep{S200224ca_update}. Due to weather conditions, these observations started with $\sim$3 hours of delay and lasted for 1.4 hr. \\

A new camera for the 28 m telescope was in commissioning phase during the S200224ca follow-up. Therefore, to have a consistent reconstruction for all observed GW events, we exclude the 28 m telescope and only use the data from only the 12 m telescopes for all GW event follow-up analyses discussed here. We note that the 28/,m telescope data were analyzed and the results are consistent with the results shown in this paper with a slight decrease in energy threshold. We also note that during the GW170814 and GW190512\_180714 observations, one of the four 12 m telescopes was not available due to maintenance. The pointing positions for the GW190728\_064510 follow up are presented in Tab.~\ref{tab:HESS_OBS1} and in Tab.~\ref{tab:HESS_OBS} of the \texttt{ADDITIONAL TABLES AND FIGURES} section for GW170814, GW190512\_180714 and S200224ca. 
Some positions were observed more than once, either due to technical or weather conditions or because there was a hint of a signal in the preliminary real-time analysis. 

\bigbreak
\section{Analysis and results}
\label{sec:analysis}

\begin{table*}[ht!]
    \centering
    \begin{tabular}{ccccccc}
    \toprule
    GW Event & Redshift & $\gamma(E = E_{\text{th}}$, $z= z_{GW})$ & Energy Range (TeV) & Coverage & $T_\mathrm{start}, T_\mathrm{stop}$ (s)\\ 
    \toprule
    GW170814  & 0.12$^{+0.03}_{-0.04}$ & 2.73 & 0.42-34.80 & 75.4\% & $2.22\times10^5, 4.10\times10^5$\\
    GW190512\_180714 & 0.27$^{+0.09}_{-0.10}$ & 3.50 & 0.31-38.31 & 34.5\% & $1.84\times10^4, 2.82\times10^4$\\
    GW190728\_064510  & 0.18$^{+0.05}_{-0.07}$ & 2.98 & 0.35-26.10 & 50.8\% & $4.88\times10^4, 7.28\times10^4$ \\
    S200224ca& 0.29 & 3.08 & 0.24-38.31 & 62.13\% & $1.07\times10^4, 1.59\times10^4$ \\
    \hline
    \end{tabular}
    \caption{Spectral indices ($\gamma$) at a GW event's corresponding redshift~\citep{Abbott_2019,abbott2020gwtc2} and at $E_{\text{th}}$ assuming an intrinsic $E^{-2}$ spectrum. The redshift for S200224ca was estimated from the distance in~\citet{S200224ca_update} using the cosmological parameters from~\citet{planck_2015}. The energy range used to derive the specific integral upper limit maps and the corresponding coverage are presented in columns 4 and 5 respectively and the last column lists the start and end of the H.E.S.S. observations of the GW event, as calculated from the reported GW merger time.}
    \label{tab:GW_REDSHIFT_INDEX_COV}
\end{table*}

\begin{figure*}
  \centering
  \begin{minipage}[b]{0.32\textwidth}
    \includegraphics[width=\textwidth]{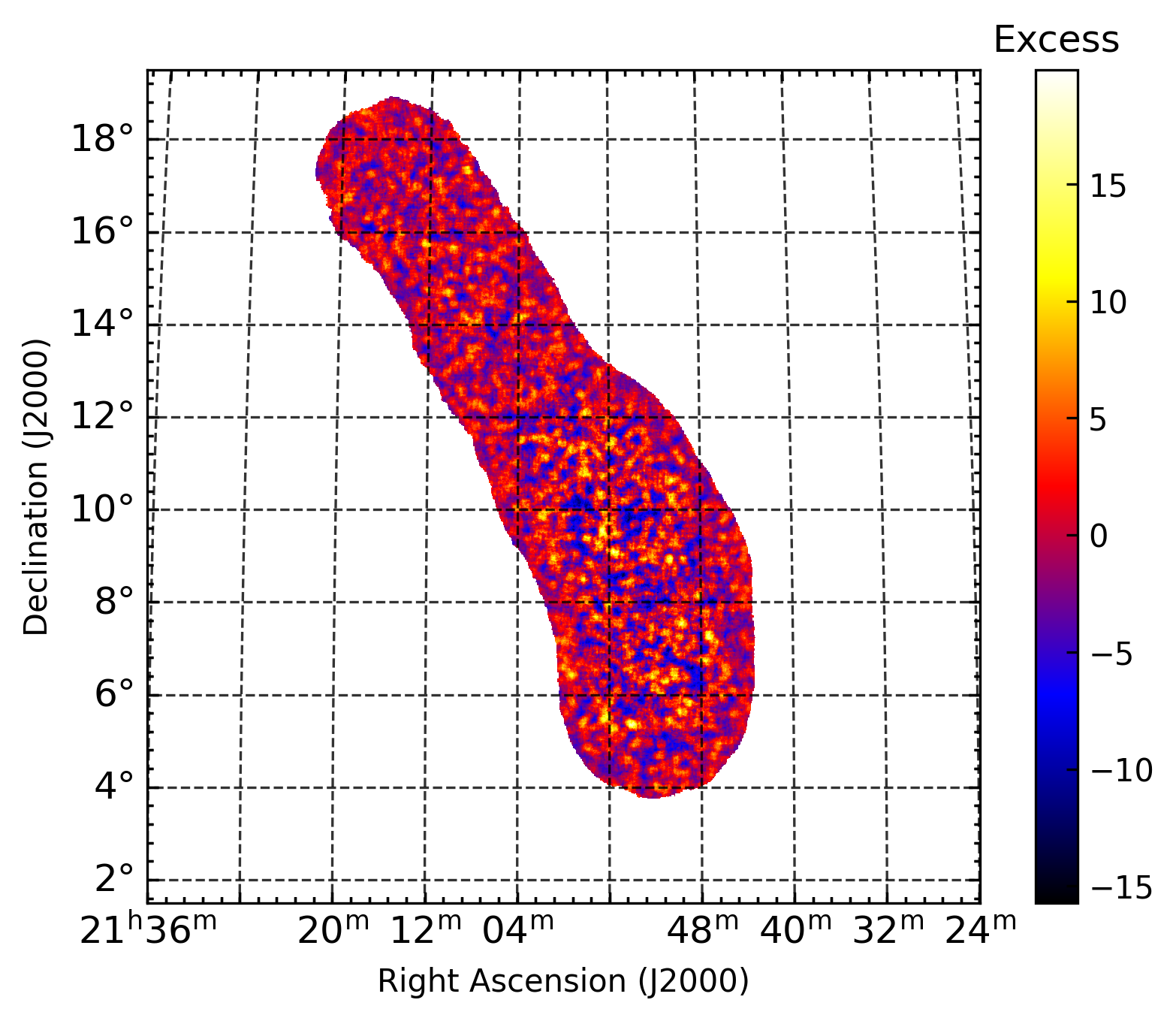}
  \end{minipage}
    \begin{minipage}[b]{0.32\textwidth}
    \includegraphics[width=\textwidth]{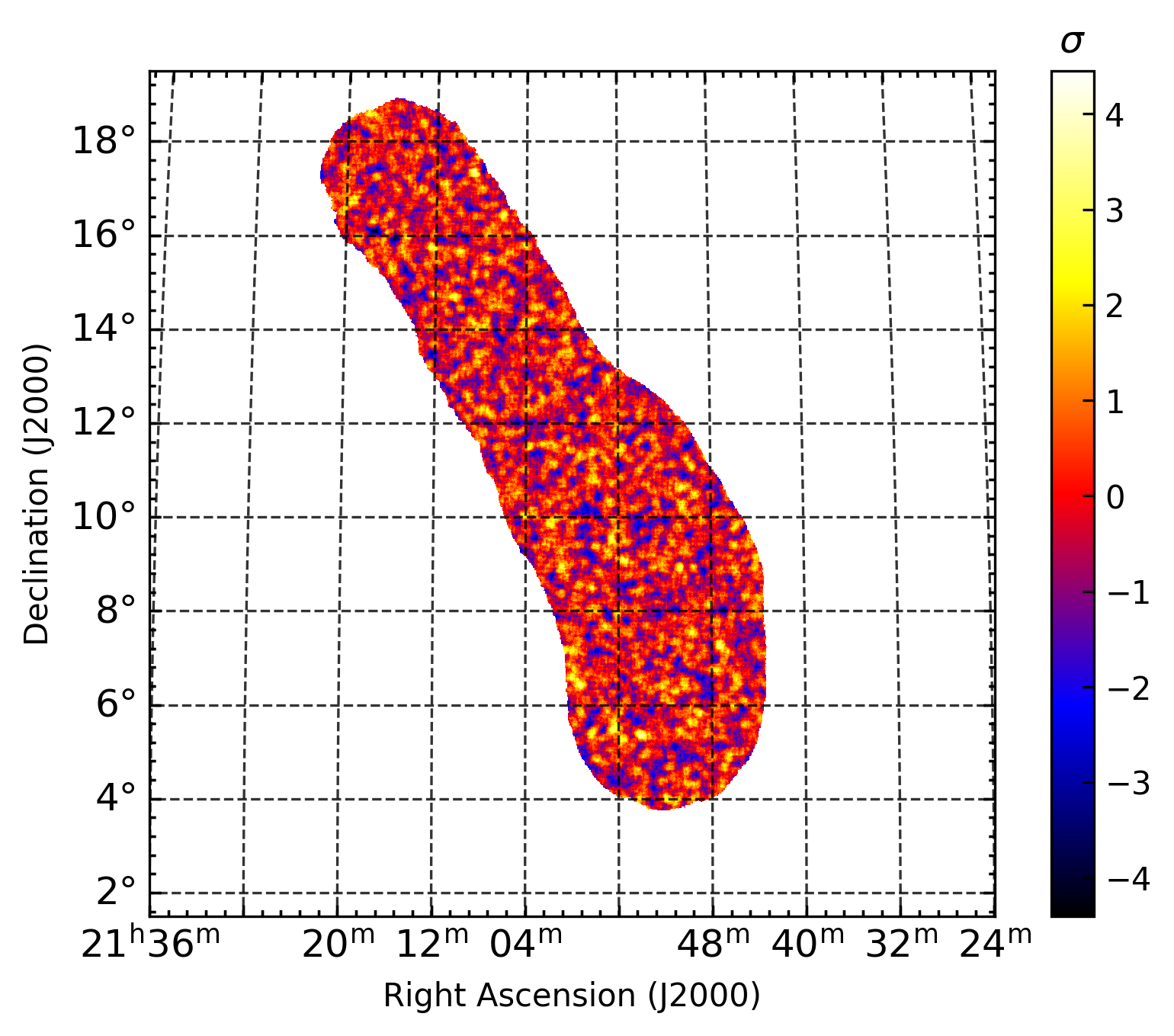}
  \end{minipage}
  \begin{minipage}[b]{0.28\textwidth}
    \includegraphics[width=\textwidth]{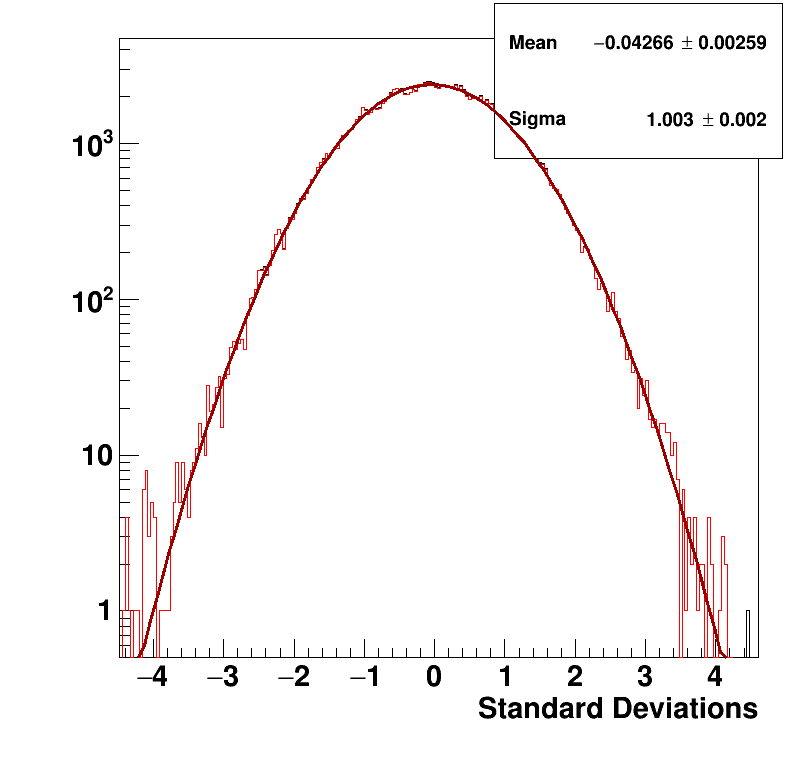}
    \end{minipage}
    \caption{Left: Excess map computed from the H.E.S.S. observational data taken on GW190728\_064510 presented in Tab.~\ref{tab:HESS_OBS1} with 0.1$^{\circ}$ oversampling radius. Middle: Significance map computed from the H.E.S.S. excess map of GW190728\_064510 . Right: Significance distribution of the H.E.S.S. significance map entries in red and a Gaussian distribution fit in black.}
    \label{fig:HESS_RingMaps}
\end{figure*}

The data analysis is performed with the Model Analysis \citep{deNaurois2009}, which is a reconstruction method that performs log-likelihood comparison between the recorded shower images and a semianalytical model of $\gamma$-ray showers. As explained in Sec.~\ref{sec:observations} the data for the four 12\,m telescopes have been analyzed in stereoscopic mode.  The background level in the FoV is determined from the data set itself using the ring background technique~\citep{Berge2007}. For that, all $\gamma$-ray-like events from the runs are accumulated, and a reconstructed map with 0.02$^{\circ}$ pixel size is created. On each position (pixel) of the map, the ON region is defined as a $0.1^\circ$ radius circular region. 
The excess counts are computed by subtracting 
the background of $\gamma$-ray-like events after analysis and selection cuts, determined from a ring region around the ON region and accounting for the radial variation of acceptance. As an example, we show in the left plot of Fig.~\ref{fig:HESS_RingMaps} the resulting excess map for the GW190728\_064510 GW event. The edges of the FoV are constrained by the statistics of the $\gamma$-ray-like background events. We only consider the regions where $\alpha  N_\text{OFF} > 5$, where $N_\text{OFF}$ is the number of events in the background region and $\alpha$ is the ratio of the effective exposure in the signal region over the effective exposure in the background region. Significance values are then computed following~\citet{LiMa}. A significance map is computed as shown in the middle panel of Fig.~\ref{fig:HESS_RingMaps}. The significance distribution shown in the right panel follows a Gaussian distribution and is consistent with the background-only hypothesis. The same is found for the other three GW events. Therefore, we conclude that the analyses resulted in no significant detection of any $\gamma$-ray counterparts to the BBH merger events.
These findings have been confirmed with an independent analysis using the Image Pixel-wise fit for Atmospheric Cherenkov Telescopes (ImPACT;~\citet{Parsons2014} software, yielding consistent results.\\


\begin{figure*}
  \centering
    \includegraphics[width=0.85\textwidth]{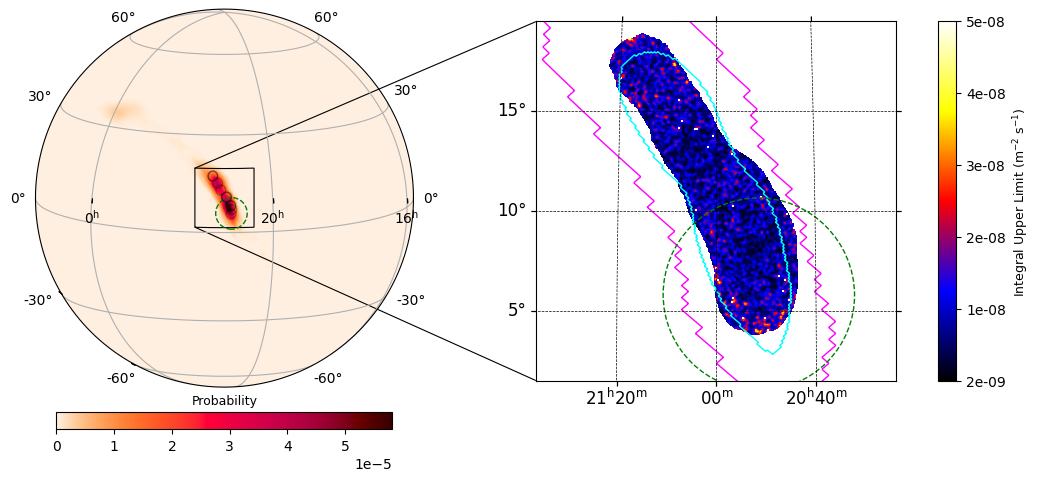}
    \caption{Left: The sky in equatorial coordinates with the probability map of the GW localization of GW190728\_064510 . Darker colors indicate regions with higher localization probabilities. Right: Integral upper-limit maps for the 1-10 TeV energy range computed from the H.E.S.S. observations of GW190728\_064510 GW events presented in Tab.~\ref{tab:HESS_OBS1} assuming an $E^{-2}$ source spectrum. The magenta and cyan lines represent, respectively, the 90\% and 50\% localization region contours of the newest published GW map in~\citet{abbott2020gwtc2}. The GW map is retrieved from~\citet{graceDB}. The green dashed circle represents the uncertainty region of the IceCube neutrino candidate~\citep{S190728q_neutrino}. The black circles represent the H.E.S.S observation FoVs. Darker colors indicate that a region was observed with more than one run.}
\label{fig:O3HESS_IntUL_1-10_S190728q}
\end{figure*}

In order to constrain the VHE $\gamma$-ray emission from these BBH mergers at the time of the H.E.S.S. observations, we derived integral upper-limit maps for each GW event for a given energy interval [$E_{\text{min}}$, $E_{\text{max}}$] and a given spectrum $\phi(E) \propto (E/E_0)^{-\Gamma}$, with $E_0$, the pivot energy, set to $1$~TeV. We first derive upper-limit maps in the 1--10 TeV energy range, in which H.E.S.S. is most sensitive, and for an $E^{-2}$ spectrum following the procedure described in~\citet{Galactic_plane_survey}. The upper-limit map for GW190728\_064510 is presented in Fig.~\ref{fig:O3HESS_IntUL_1-10_S190728q} and Fig.~\ref{fig:O3HESS_IntUL_1-10} of the \texttt{ADDITIONAL TABLES AND FIGURES} section presents the three remaining GW events.\\ 

We also exploit the widest energy range possible for each observation between $E_{\text{min}}$ which is the threshold energy $E_{\text{th}}$ and a maximum energy $E_{\text{max}}$. The energy range is imposed by instrument limitations and varies between runs, as it depends on the zenith angle of the position studied in the sky at the time of observation. Higher zenith angles lead to higher energy thresholds. The energy threshold $E_{\text{th}}$ and the maximum energy $E_{\text{max}}$ are chosen where the energy reconstruction bias is less than $10\%$. The $E_{\text{th}}$ is increased if it is below the energy at which the effective area is at $10\%$ of its maximum value. For the latter, a per-pixel effective area constrained energy threshold is considered, and a median over the entire coverage is taken. The resulting energy ranges are presented in Tab.~\ref{tab:GW_REDSHIFT_INDEX_COV}.


For the intrinsic spectrum shape, the same generic $E^{-2}$ power-law spectrum is assumed. Extragalactic background light (EBL) in the ultraviolet, optical, and infrared bands interacts with VHE $\gamma$-rays via electron-positron pair production absorbing parts of the VHE $\gamma$-ray flux. The EBL absorption effects increase with energy. Given the EBL absorption~\citep{Franceschini} at the respective redshift of each GW event, we evaluate the equivalent power law at the threshold energy $E_{\text{th}}$ of each GW event and compute the equivalent spectral indices. The $E_{\text{th}}$ is considered due to the fact that low-energy events are expected to be dominant. The attenuation due to EBL causes the observed spectrum to be softer than the intrinsic one. The resulting power-law indices are presented in Tab.~\ref{tab:GW_REDSHIFT_INDEX_COV} and the effective power-law spectra are used to compute specific upper-limit maps. Varying the intrinsic spectral index between 1.5 and 2.5 leads to a variation of the equivalent spectral index between +16\% and -16\% for the absorbed emission. This affects the integral upper-limit values between +19.4\% and -10.7\%. Moreover, the equivalent spectral index shows less than 14\% dependence on the EBL model in the lower part of the energy spectrum up to 1 TeV. Propagating this dependency to the upper-limit maps will result in less than 10\% variation in the upper-limit values.

The 95\% C.L. integral upper limits are computed with the Rolke method~\citep{Rolke} at each observed position of the map for all four BBH mergers within the energy ranges defined in Tab.~\ref{tab:GW_REDSHIFT_INDEX_COV} as explained above using the effective power-law spectra. The maps are presented in Fig.~\ref{fig:O3HESS_IntUL_Spe} in the \texttt{ADDITIONAL TABLES AND FIGURES} section for all four GW events. The effective VHE $\gamma$-ray coverage of the GW localization region after data selection and analysis is presented in the last column of Tab.~\ref{tab:GW_REDSHIFT_INDEX_COV}. 

\bigbreak
\section{Discussion}
\label{sec:discussion}

The H.E.S.S. observations presented in this paper cover large portions of the GW localization regions, as shown in Tab.~\ref{tab:GW_REDSHIFT_INDEX_COV}. The computed upper limits constrain the VHE $\gamma$-ray emission from BBH mergers for the first time. No electromagnetic or neutrino counterpart has been confirmed for any of the GW events observed by H.E.S.S. \\

The models that predict electromagnetic emission from BBH mergers are primarily inspired by the candidate electromagnetic counterpart to GW150914. 
Because these models focus on the production of X-ray and low-energy $\gamma$-ray emission, they do not make many statements about VHE emission. For instance, the scenarios involving a remnant disk make predictions on the electromagnetic emission based on synchrotron radiation from the relativistic jets that are launched after the merger. Assuming that the particle acceleration and radiation of the parent charged particle population occur in the same region, there is a maximum synchrotron photon energy that can be estimated by equating the particle acceleration and emission timescales; for GRB-like conditions, after a few hours, this tends to be $\mathcal{O}$(1 GeV) \citep[e.g.,][]{GRB190829A_paper}, more than an order of magnitude below the H.E.S.S. energy range and 2 orders of magnitude below the GW event-specific energy ranges discussed here (Tab.~\ref{tab:GW_REDSHIFT_INDEX_COV}). The H.E.S.S. upper limits therefore do not constrain these models, and an additional photon production mechanism --- such as inverse Compton emission --- would be necessary to generate a potentially detectable VHE signal. Note that, while VHE emission from GRBs has been detected, its origin is not completely understood, and its applicability to BBH mergers is unclear.
Given the amount of uncertainty involved, any predictions on the production of VHE emission from BBH mergers hours after the merger event is beyond the scope of this paper. \\

While the VHE emission from a BBH merger is difficult to predict, we can still compare the energy flux and luminosity upper limits for the BBH events to the measurements of GRBs, given the proposed models that link GRBs and BBH mergers. 
The energy flux upper limits for the H.E.S.S. GW observations were calculated both with and without correcting for EBL absorption. The EBL-corrected energy flux upper limits $F_\mathrm{unabs}$ are used for calculating the luminosity upper limits and are calculated assuming an intrinsic $E^{-2}$ spectrum over 1--10 TeV in the source frame. This results in an event-specific energy range of $\frac{1}{1+z}$ to $\frac{10}{1+z}$~TeV in the observer frame. Note that calculating the intrinsic values in this way in general runs the risk of integrating outside of an instrument's bandpass, as the energy range over which spectral fitting is performed is usually defined by instrumental limitations; however, $\frac{1}{1+z}$ to $\frac{10}{1+z}$~TeV is safely within the event-specific energy ranges (Tab.~\ref{tab:GW_REDSHIFT_INDEX_COV}) for all of the events discussed here. 

\begin{figure*}
  \centering
    \includegraphics[width=0.7\textwidth]{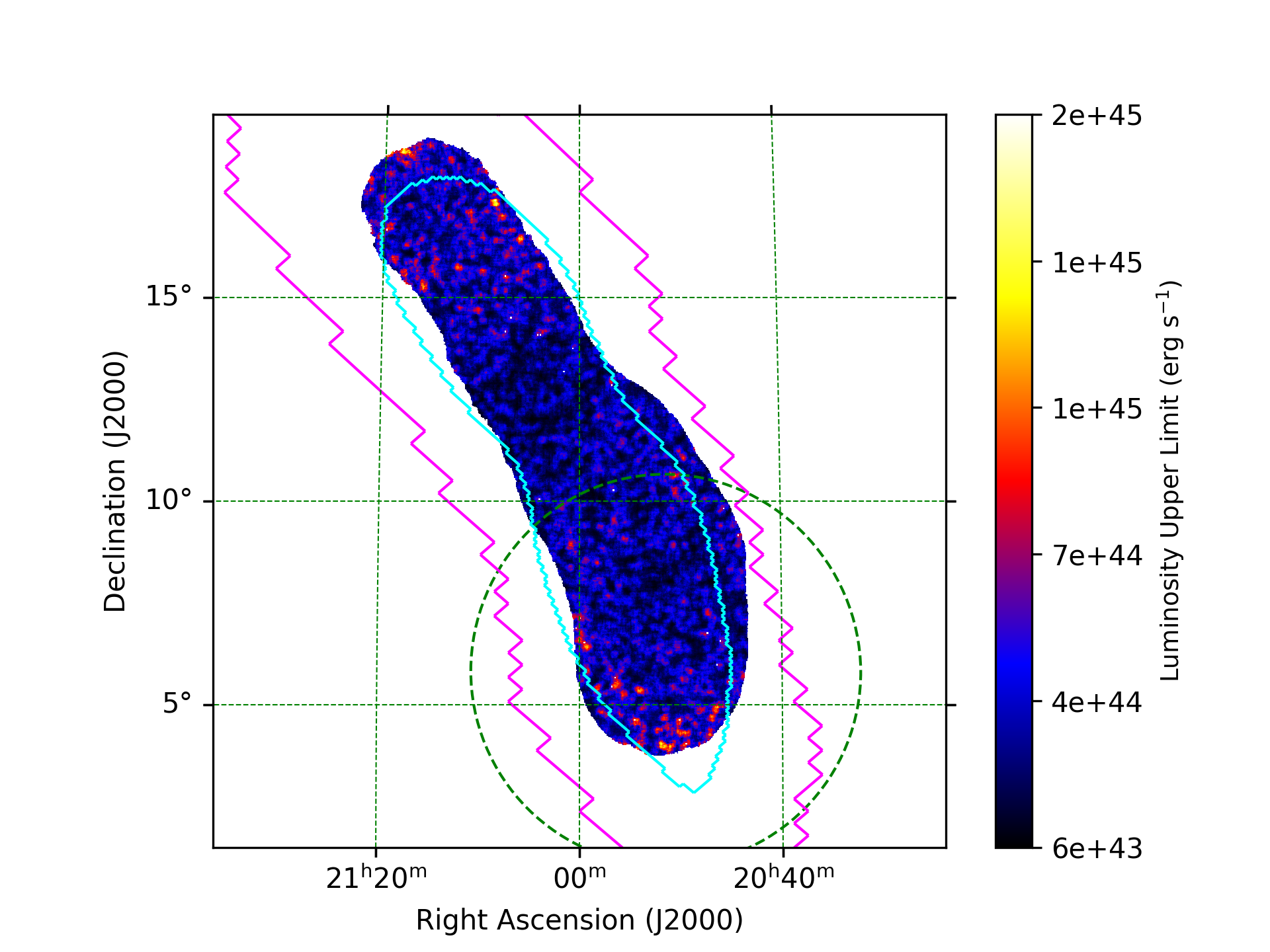}
    \caption{Luminosity upper-limit map computed from the H.E.S.S. upper-limit map for GW190728\_064510 presented in Fig.~\ref{fig:O3HESS_IntUL_1-10_S190728q}}
\label{fig:O3HESS_Lum_S190728q}
\end{figure*}

We then calculate the isotropic luminosity from the unabsorbed energy flux. In short,
\begin{align}
  L_\mathrm{iso} = 4\pi D_\mathrm{L}^2 \int_{E_1/(1+z)}^{E_2/(1+z)} E N_0 \left(\frac{E}{E_0}\right)^{-\Gamma} dE
\end{align}
where $E_1=1$~TeV, $E_2=10$~TeV, $\Gamma=2$, and $D_L$ is the luminosity distance. (Here $N_0$ encompasses the observation-specific measurements; see Eqn.\,2 of \citet{Galactic_plane_survey}.) We use the per-pixel luminosity distance from the latest published GW sky maps as, for a given GW event, the mean luminosity distance can vary by several hundred megaparsecs in the regions observed by H.E.S.S. 
The EBL-absorbed energy flux upper limits are calculated with the GW event-specific spectra and energy ranges (Tab.~\ref{tab:GW_REDSHIFT_INDEX_COV}). The luminosity upper-limit map for GW190728\_064510 is shown in Fig.~\ref{fig:O3HESS_Lum_S190728q} and the maps for the three remaining events are shown in Fig.~\ref{fig:O3HESS_Lum_all} of the \texttt{ADDITIONAL TABLES AND FIGURES} section.\\ 


\begin{table*}[ht!]
    \centering
    \begin{tabular}{ccccccc}
    \toprule
    GW Event & & \multicolumn{2}{c}{Energy Flux, Event-specific (erg cm$^{-2}$ s$^{-1}$)} & & \multicolumn{2}{c}{Luminosity, Standard (erg s$^{-1}$)}\\\cline{3-4}\cline{6-7}
    & & Mean & Standard Dev. & & Mean & Standard Dev. \\
    \toprule
    GW170814 && $3.7\times10^{-12}$ & $1.8\times10^{-12}$ & & $1.3\times10^{44}$ & $9.8\times10^{43}$\\
    GW190512\_180714 && $3.1\times10^{-12}$ & $1.5\times10^{-12}$ & &$9.9\times10^{44}$ & $4.7\times10^{44}$\\
    GW190728\_064510 && $2.6\times10^{-12}$ & $1.3\times10^{-12}$ & &$3.2\times10^{44}$ & $1.6\times10^{44}$\\
    S200224ca&& $2.7\times10^{-12}$ & $1.2\times10^{-12}$ & &$1.9\times10^{45}$ & $8.8\times10^{44}$\\
    \hline
    \end{tabular}
    \caption{The energy flux and luminosity upper limits for the four GW events were calculated individually for each pixel in the sky region observed by H.E.S.S. The first column lists the GW event discussed in this paper. The second two columns list the mean and standard deviation of the GW event-specific energy flux upper limits, not corrected for EBL absorption, and calculated with the event-specific energy ranges and indices (Tab.~\ref{tab:GW_REDSHIFT_INDEX_COV}). The third and fourth columns list the mean and standard deviation of the upper limits on isotropic luminosity, calculated from the EBL-corrected energy fluxes assuming an $E^{-2}$ source spectrum over a 1--10 TeV energy range, and using the per-pixel luminosity distances. These values are plotted in Figs.~\ref{fig:luminosity_comparisons} and \ref{fig:flux_comparisons}.}
    \label{tab:flux_lum_ULs}
\end{table*}

Among the GRBs that have been detected by VHE telescopes, the two detected by H.E.S.S. \citep{GRB180720B,GRB190829A_paper} have measurements hours after the GRB began, which is roughly the same time delay as for the GW events, so we include these two in our comparison sample. In order to enlarge the sample, we also take the set of GRBs with known redshift and temporally extended emission observed by the Large Area Telescope (LAT) on board the Fermi space observatory~\citep{LAT_GRBs}. For the luminosity estimates, we consider the energy flux at late times measured by the LAT in the 100~MeV -- 100~GeV energy range. Using the spectral index $\gamma_\mathrm{LAT}$ measured by the LAT at these times, we extrapolate the spectrum to calculate the energy flux in the H.E.S.S. energy band, assuming an optimistic scenario in which the index stays the same. We then extrapolate these forward in time using the late-time power-law decay index measured by the LAT and convert the energy fluxes into isotropic luminosities using the redshifts of the GRBs.

We emphasize that these extrapolations should be considered more as guidelines to illustrate a range of potential behavior and should not be considered as strict predictions of VHE emission from BBH mergers. While short GRBs are the ones that are associated with binary mergers, all but one of the LAT GRBs (and both of the H.E.S.S. GRBs) in our sample are long GRBs, which have been observed to have isotropic energy releases that are, on average, 1 or 2 orders of magnitude larger than those of short GRBs \citep{Li_2016}. In addition, the extrapolations themselves are rather simplistic and result in overpredictions on the VHE flux. Some of the LAT GRBs have spectral indices of around 2, suggesting that the spectrum must soften somewhere above the LAT energy range --- potentially in the H.E.S.S. energy range --- to avoid the energy flux diverging, whereas we have assumed that it continues unchanged. In addition, the spectral index should evolve with time as the blast wave decelerates and the densities decrease, whereas we have not included any spectral evolution. To estimate the effect of our assumption and illustrate the magnitude of uncertainty, we also calculate what the energy flux will be if the spectral index softens when going from the LAT to the H.E.S.S. energy range. We find that if the spectrum softens from $\gamma_\mathrm{LAT} \to \gamma_\mathrm{LAT} + 0.5$, the luminosity drops by a factor of around 50; if the spectrum softens to $\gamma_\mathrm{LAT} + 1$, the luminosity drops by three orders of magnitude. \\

\begin{figure*}[ht!]
  \centering
    \includegraphics[width=0.7\textwidth]{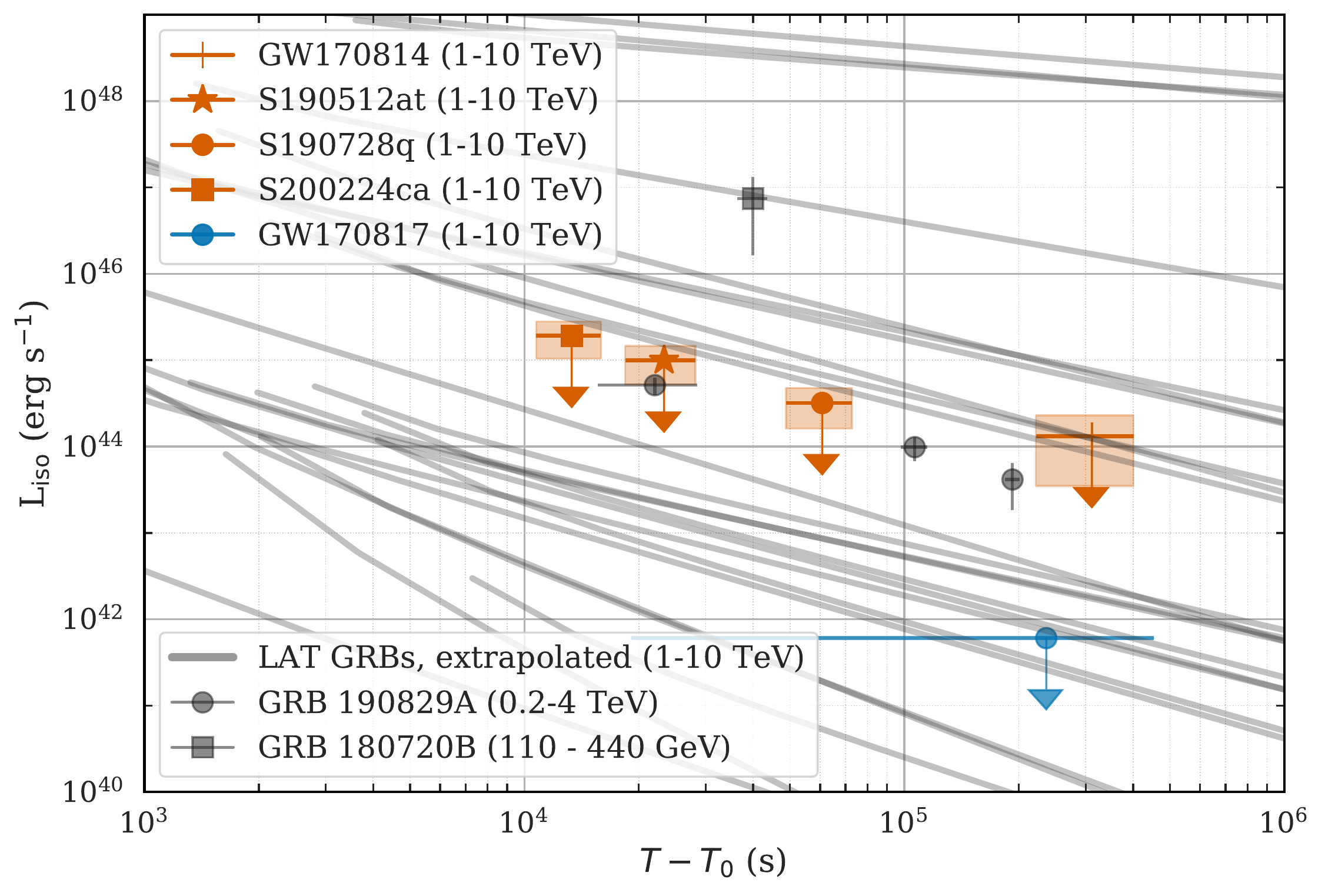}
    \caption{The upper limits on isotropic luminosity, $L_\mathrm{iso}$, and the observation time as measured from the merger, $T - T_0$, are shown for the four BBH events. 
    For each of these events, we present the mean (orange points and arrows) and the standard deviation (orange bands) of the per-pixel luminosity upper-limit maps (e.g., Fig.~\ref{fig:O3HESS_Lum_S190728q}). We also include the upper limit for GW170817 measured by H.E.S.S. (\citet{HESS170817_deep}; blue point and arrow). For comparison, we also include the luminosity measurements of the H.E.S.S.-detected GRBs 190829A (gray circles) and 180720B (gray square) and extrapolations of LAT GRBs with measured redshifts and temporally extended emission (gray lines). A subset of the LAT GRB extrapolations is not visible below the y-axis minimum.}
\label{fig:luminosity_comparisons}
\end{figure*}

The luminosity comparisons are shown in Fig.~\ref{fig:luminosity_comparisons}. The upper limits for the four BBH events lie below some of the extrapolations of LAT GRBs --- given that the GRBs in general are located at larger distances --- and are at a similar level as the VHE-detected GRB 190829A, which is at a similarly low redshift. When comparing the four BBH events, it can be seen that the two closer events (GW170814 and GW190728\_064510 ) have the lowest luminosity upper limits. The value of the upper limit is not greatly affected by the observation duration; since the observations are tiled, increasing the duration increases the amount of sky coverage rather than the sensitivity to a signal in any particular part of the sky. In comparison, the upper limit for GW170817 --- taken over the entire observation duration - is 3 orders of magnitude lower, primarily because of this event's proximity but also due to the deeper observations of this well-localized source \citep{HESS170817_deep}. \\

By comparing the luminosities, we are comparing the intrinsic properties of the source. In order to compare the sources as they are observed on Earth, we additionally calculate the energy flux upper limits using the EBL-attenuated spectra with the power-law index and energy range given in Table~\ref{tab:GW_REDSHIFT_INDEX_COV}. For the VHE-detected GRB 180720B, we calculate the energy flux for the observed spectrum using the simple power-law fit to the data (i.e., the fit that does not correct for the EBL absorption). For GRB 190829A, a power-law fit to the data is not provided for the third night on its own, so we calculate the EBL-attenuated energy flux for each of the 3 nights separately using the constant intrinsic photon index of 2.07 derived by combining the data from all three nights. For the LAT GRBs, we take the extrapolated spectra and calculate the effect of the EBL on these spectra if the LAT GRBs were all at the redshift of GW190728\_064510 , over the specific energy range of GW190728\_064510 . (Using one of the other BBH events results in a decrease in the energy flux extrapolations by at most 50\% for GW190512\_180714 and S200224ca and an increase of less than 75\% for GW170814.)\\

\begin{figure*}[!ht]
  \centering
    \includegraphics[width=0.7\textwidth]{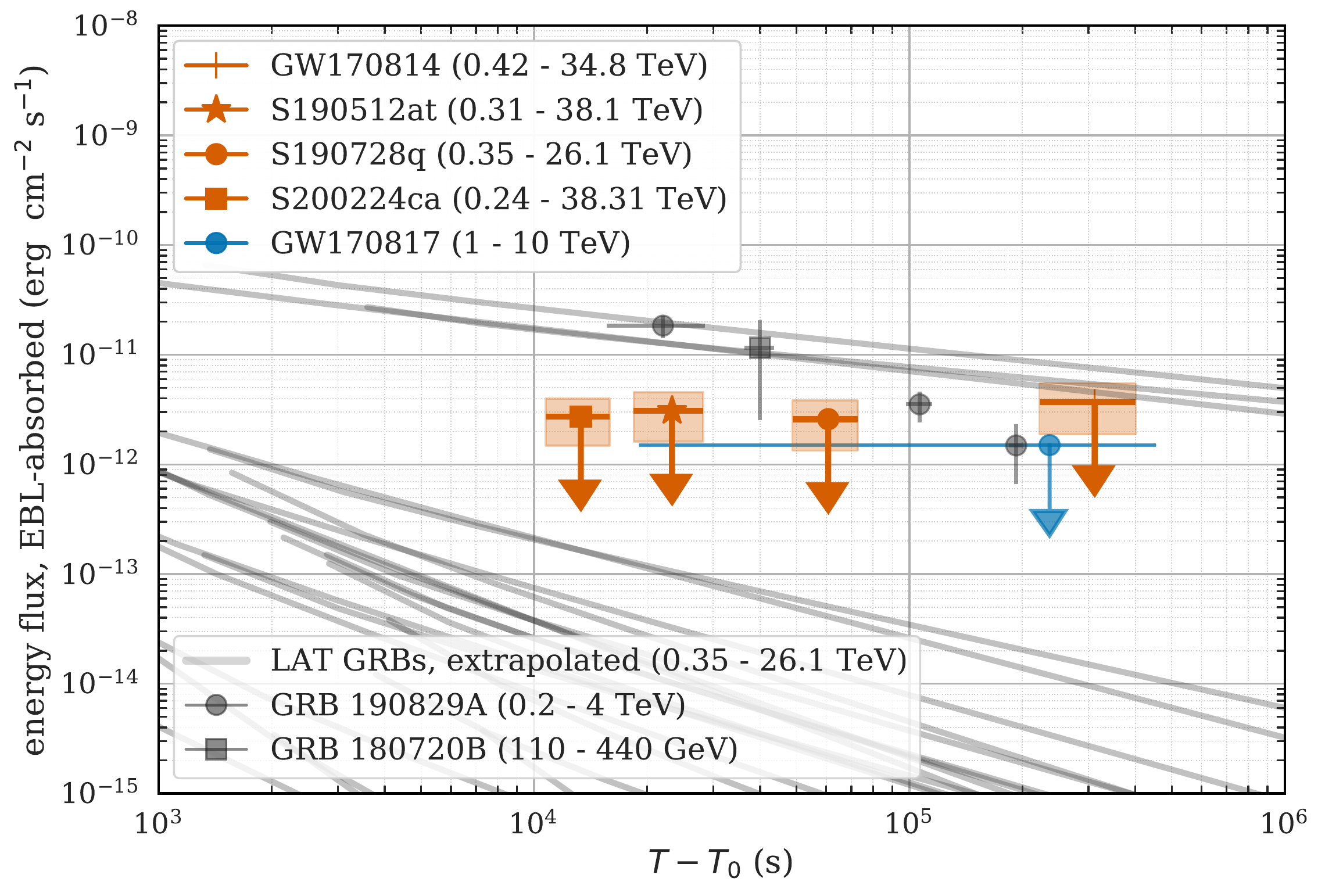}
    \caption{The upper limits on the observed energy flux and the observation time as measured from the merger, $T - T_0$, are shown for the four BBH events. For each event, we present the mean (orange points and arrows) and standard deviation (orange bands) on the integral energy flux upper limits calculated from the per-pixel EBL-absorbed integrated energy flux upper-limit maps (available at \url{https://www.mpi-hd.mpg.de/hfm/HESS/pages/publications/auxiliary/2021_BBH_O2_O3/}). We also include the upper limit for GW 170817 as measured by H.E.S.S. (\citet{HESS170817_deep}; blue point and arrow). For comparison, we also include the energy flux measurements of GRBs 190829A (gray circles) and 180720B (gray squares), as well as the extrapolations of LAT GRBs (gray lines), calculated using the specific energy band and redshift of GW190728\_064510 . A subset of the LAT GRB extrapolations is not visible below the y-axis minimum.}
    \label{fig:flux_comparisons}
\end{figure*}

The average energy flux upper limits for the GW events are shown in Fig.~\ref{fig:flux_comparisons}, along with the LAT GRB extrapolations and the measured values for the H.E.S.S.-detected GRBs. The third night of GRB 190829A (gray circles) resulted in an energy flux measurement below the GW upper limits, mainly because the GRB measurement was the result of almost 5 hr of observations on a single position. (The difference in the energy ranges does not greatly affect this comparison; assuming the measured GRB 190829A spectrum extends up to 40 TeV increases the absorbed energy flux by 7\%, and the increase is negligible for GRB 180720B given its larger distance.) In contrast, for GW observations, no single sky position gets as much exposure as it would in a standard single-position multihour GRB follow-up.

The upper limits on the energy flux of the BBH events lie over an order of magnitude above almost all of the LAT GRB extrapolations. In addition, the upper limits for the four events are found at similar levels, although they span a large range in observation delays. In order to test whether BBH mergers produce gamma-ray radiation in the H.E.S.S. energy range at a level similar to these extrapolations of LAT GRBs, the H.E.S.S. upper limits would need to be comparable to these extrapolations. This indicates that either shorter observation delays or deeper, more sensitive observations by H.E.S.S. are required to test this. \\

The temporal extrapolations of the LAT GRBs in Figs.~\ref{fig:luminosity_comparisons} and \ref{fig:flux_comparisons} assume that these are all on-axis events. Indeed, this assumption is valid, given that all of the GRBs in this group had afterglow light curves that decayed with time, whereas off-axis events such as GRB 170817A produce emission that rises at early times \citep{Troja2020_170817xrayAG, Nynka2018_170817xrayAG}. Given the current typical sensitivities of GW detectors, the inclination angles --- i.e., the angle between the line of sight and the total angular momentum vector --- of most GW events can only be weakly constrained due to the degeneracy between the inclination angle and the source distance \citep{Nissanke_2010,Usman_2019}. It can only be constrained by knowledge of the host galaxy, which is difficult to ascertain without an electromagnetic counterpart. Because of this, the inclination angles of the events discussed here are mostly unconstrained and so cannot be used to further predict the behavior of any potential electromagnetic emission.


\section{Outlook and conclusion}
\label{sec:outlook}

During O2 and O3, H.E.S.S. observed four BBH mergers: GW170814, GW190512\_180714, GW190728\_064510, and S200224ca. 
For these GW events, the delay between merger and H.E.S.S. observation is at least a few hours, due to the necessity of waiting for the part of the sky containing the merger to be at favorable zenith angles above the telescopes. 
The low rate of sufficiently well-localized GW events in the previous Advanced LIGO and Advanced Virgo runs meant we were unlikely to have observed an event with a favorable time and position.
However, this is expected to change in O4 and beyond. By comparing the expectations for O3 and O4~\citep{LVK_LivingReviews},\footnote{We note that the merger rates of GW events are updated regularly (e.g., \citet{abbott2020gwtc2}), and the numbers discussed here should only be considered as illustrative.} we can see that, prior to the start of O3, only around 25\% of GW events in a three-interferometer scenario were expected to have 90\% credible regions (high-latency off-line analysis), similar to the values for the four events discussed here (90\% credible region $\lessapprox100$ deg$^2$); from \citet{graceDB}, it can be seen that this was an accurate prediction of the low-latency localizations in O3. For an event rate of $18^{+53}_{-12}$ yr$^{-1}$ (combining binary neutron stars, neutron star-black hole, and BBH events for simplicity), this translates to only a handful of events per year with sufficiently small credible regions to pass the requirements for H.E.S.S follow up. This number increases to 75\% of events for the four-interferometer network in O4, and with an estimated event rate of $90^{+232}_{-55}$ per year, this means around $67^{+174}_{-41}$ mergers yr$^{-1}$ would pass the requirements for H.E.S.S. follow-up. Given that around half the sky is observable by H.E.S.S., and taking into account that, on average, H.E.S.S. can observe for $\sim\!6$ hr a night (averaging over a year), this results in approximately $8^{+22}_{-5}$ events yr$^{-1}$ occurring during darktime, i.e., being observable with minimal latency. For well-localized GW events, H.E.S.S. observations can begin within 1 minute. On the GW side, in O3, the preliminary notices arrived with a latency of $\approx 10$~minutes \citep{graceDB}; for O4, this would mean the upper limits in Figs.~\ref{fig:luminosity_comparisons} and \ref{fig:flux_comparisons} moving to values of $T - T_0 < 1000$~s for around eight events yr$^{-1}$ (not accounting for bad weather). For a fraction of GW events, this latency will become negligible or even negative (as measured from the merger time) as GW detector sensitivity improves with the implementation and improvement of the early-warning system \citep{Nitz_2020_earlywarning,Sachdev_2020_earlywarning, Magee_2021_earlywarning}.

Alternatively, given the more precise localizations expected in O4 and beyond, H.E.S.S. could instead choose to spend time on deeper observations of well-localized GW events. In O4, around 35\% of mergers observed by the four-interferometer network are expected to have 90\% credible areas of less than 20 deg$^2$, which would translate to 50\% credible areas of less than a few square degrees. For the expected total merger rate of $90^{+232}_{-55}$ yr$^{-1}$ , given that half of these events will be too far north, and roughly half of them will be behind the Sun, this translates to around $8^{+20}_{-5}$ events yr$^{-1}$  for which the 50\% credible region could be covered with only one pointing within 24 hr. This would then allow the energy flux upper limit in Figure~\ref{fig:flux_comparisons} to improve by up to a factor of $\sim\!5$, depending on the amount of observation time, given that the maximum amount of contiguously available observation time is around 12 hr.






Our observation strategy aims to scan over the GW localization regions, thereby emphasizing sky coverage over sensitivity. Improving the chances of a VHE detection would require reducing the delay of follow-up observations and/or spending more time observing single sky regions. Assuming that the GW events are observed fairly on-axis (i.e., that the temporal decay of any electromagnetic emission fades like GRB afterglows), we find that minimizing the observation delay (i.e., following up GW events that are immediately observable by H.E.S.S.) would have a greater effect on the detectability than reducing the sky coverage and spending more time observing single positions. This is expected to happen naturally in the next observing runs; with the increased rate of GW detections, the number of events with favorable conditions for prompt follow-up observations will also increase. Therefore, we do not expect to fundamentally alter our observing strategy and will continue to prioritize sky coverage.

We provide our upper-limit maps in fits format at \url{https://www.mpi-hd.mpg.de/hfm/HESS/pages/publications/auxiliary/2021_BBH_O2_O3/}.



\section*{Acknowledgments}
The support of the Namibian authorities and of the University of Namibia in facilitating the construction and operation of H.E.S.S. is gratefully acknowledged, as is the support by the German Ministry for Education and Research (BMBF), the Max Planck Society, the German Research Foundation (DFG),  the  Helmholtz Association,  the Alexander  von  Humboldt Foundation,  the  French  Ministry  of  Higher  Education,  Research  and  Innovation,  the  Centre  National  de  la  Recherche  Scientifique  (CNRS/IN2P3  and CNRS/INSU), the Commissariat a l'energie atomique et aux energies alternatives (CEA), the U.K. Science and Technology Facilities Council (STFC), the Knut and Alice Wallenberg Foundation,  the National Science Centre,  Poland grant No. 2016/22/M/ST9/00382, the South African Department of Science and Technology and National Research Foundation, the University of Namibia, the National Commission on Research, Science \& Technology of Namibia (NCRST), the Austrian Federal Ministry of Education, Science and Research and the Austrian Science Fund (FWF), the Australian Research Council (ARC), the Japan Society for the Promotion of Science, and by the University of Amsterdam.  We appreciate the excellent work of the technical support staff in Berlin, Zeuthen, Heidelberg, Palaiseau, Paris, Saclay, T\"ubingen, and in Namibia in the construction and operation of the equipment.  This work benefited from services provided by the H.E.S.S. Virtual Organisation, supported by the national resource providers of the EGI Federation.

\bibliographystyle{aa}
\bibliography{main}


\newpage
\section*{Additional tables and figures}
\begin{figure*}
  \centering
  \begin{minipage}[b]{0.75\textwidth}
    \includegraphics[width=\textwidth]{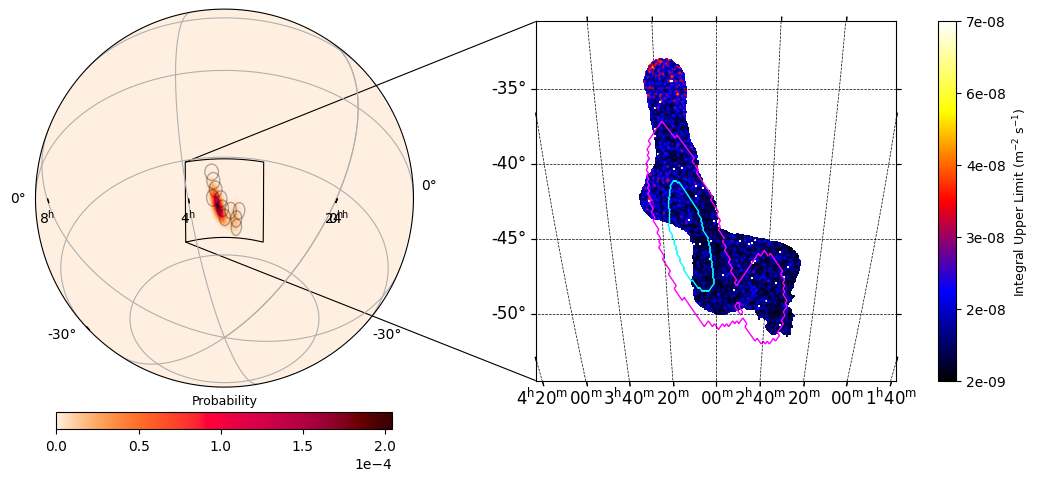}
  \end{minipage}
    \begin{minipage}[b]{0.75\textwidth}
    \includegraphics[width=\textwidth]{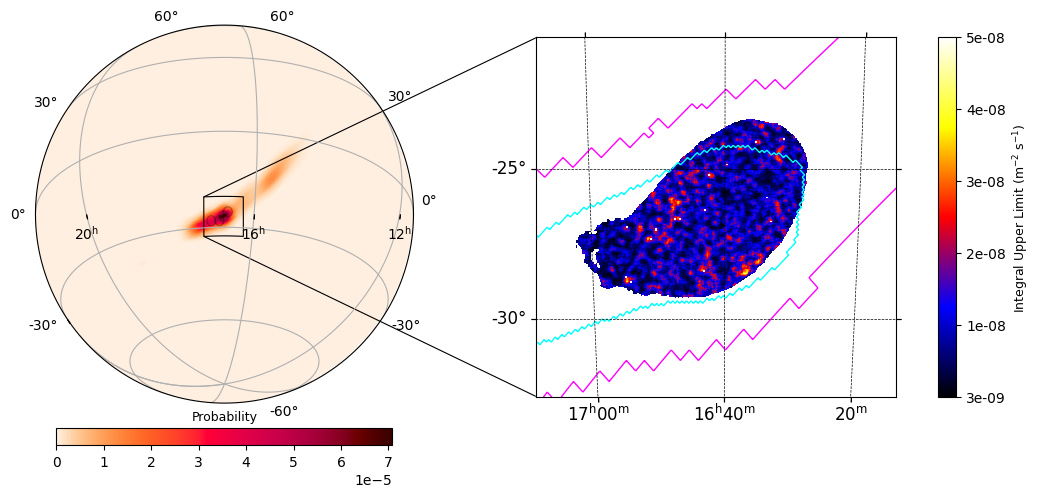}
  \end{minipage}
     \begin{minipage}[b]{0.75\textwidth}
    \includegraphics[width=\textwidth]{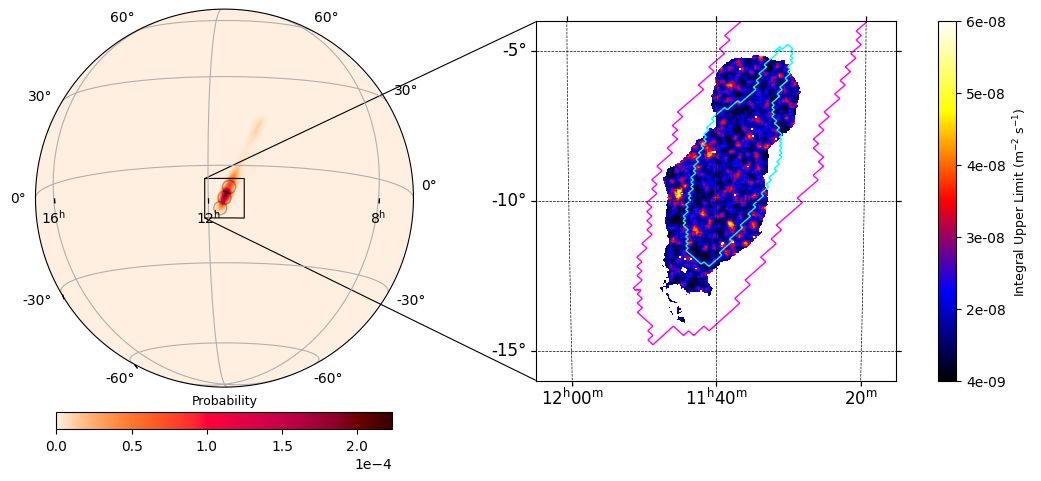}
    \caption{Left: The sky in equatorial coordinates with the probability map of the GW localizations of GW170814 (upper), GW190512\_180714 (middle), and S200224ca (lower). Darker colors indicate regions with higher localization probabilities. Right: Integral upper limit maps computed from the H.E.S.S. observations of GW170814 (upper), GW190512\_180714 (middle) and S200224ca (lower) BBH events presented in Tab.~\ref{tab:HESS_OBS} assuming an $E^{-2}$ source spectrum and a 1-10 TeV energy range. The magenta and cyan lines represent respectively the 90\% and 50\% localization contours of the newest published GW map in~\citet{Abbott_2019} for GW170814, in~\citet{abbott2020gwtc2} for GW190512\_180714 and in~\citet{S200224ca_update} for S200224ca. The GW maps are retrieved from~\citet{graceDB}. The black circles represent the H.E.S.S observation FoVs. Darker colors indicate that a region was observed with more than 1 run.}
    \label{fig:O3HESS_IntUL_1-10}
  \end{minipage}
\end{figure*}

\begin{figure*}
  \centering
  \begin{minipage}[b]{0.49\textwidth}
    \includegraphics[width=\textwidth]{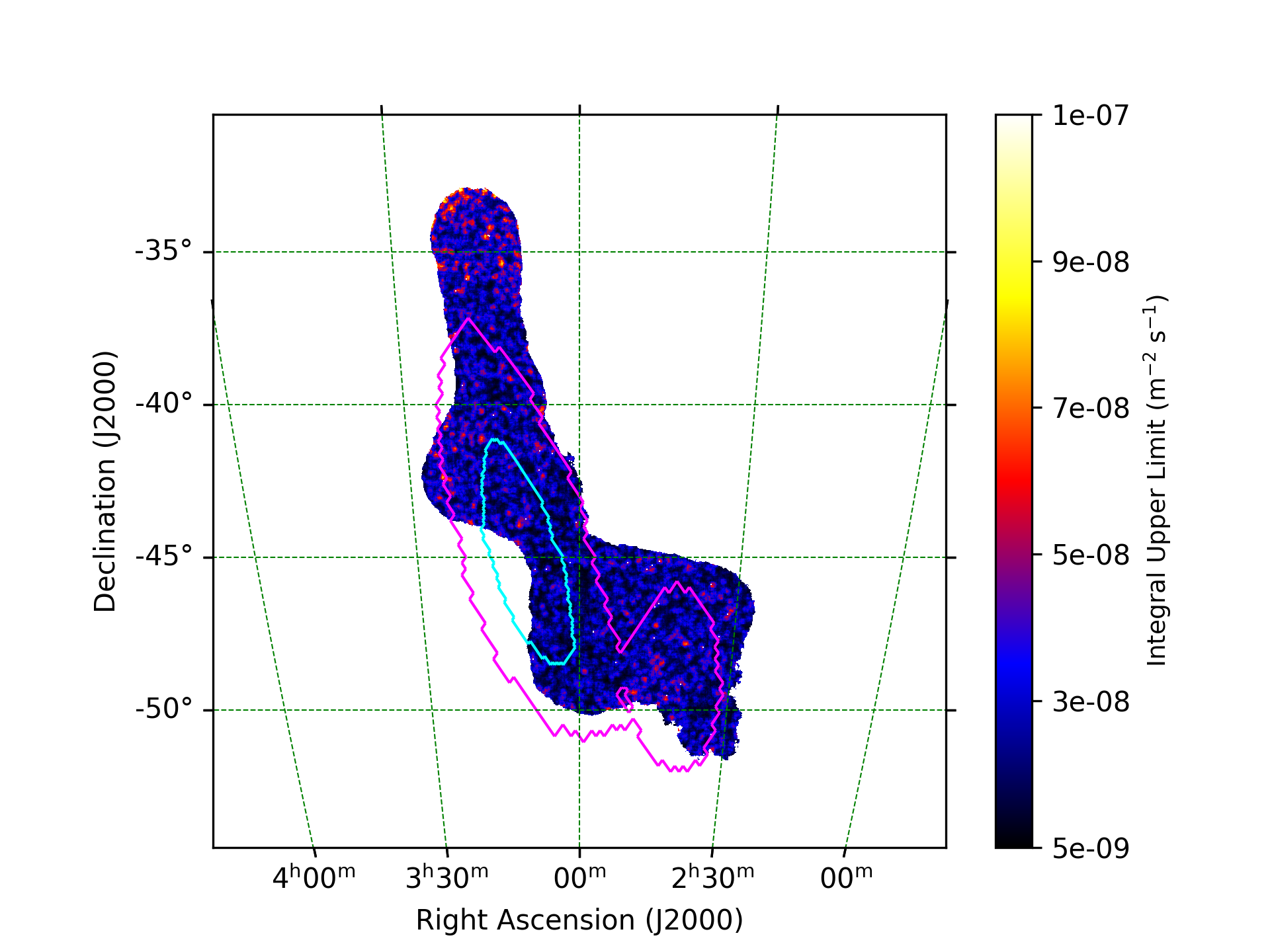}
  \end{minipage}
    \begin{minipage}[b]{0.49\textwidth}
    \includegraphics[width=\textwidth]{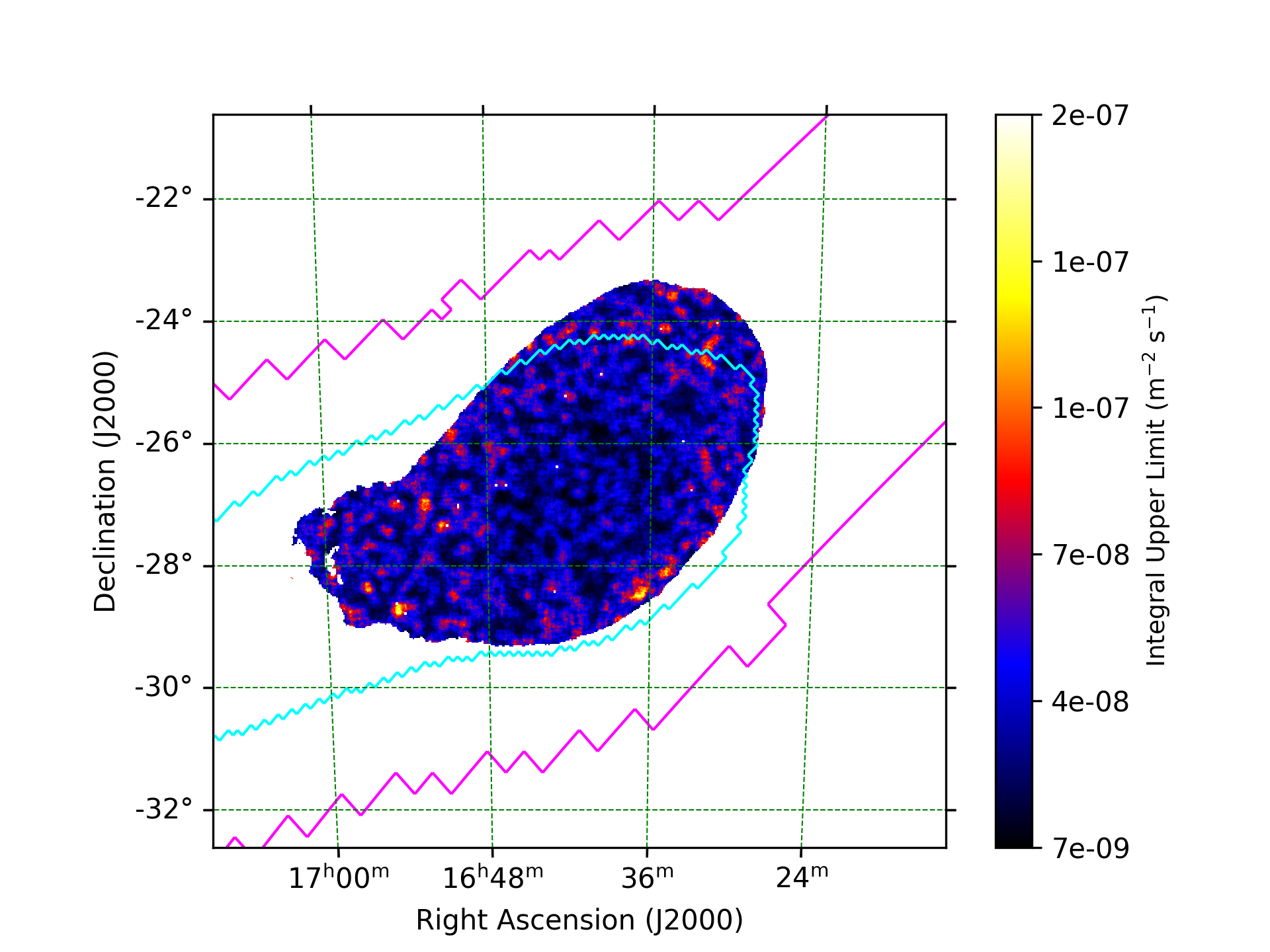}
  \end{minipage}
    \begin{minipage}[b]{0.49\textwidth}
    \includegraphics[width=\textwidth]{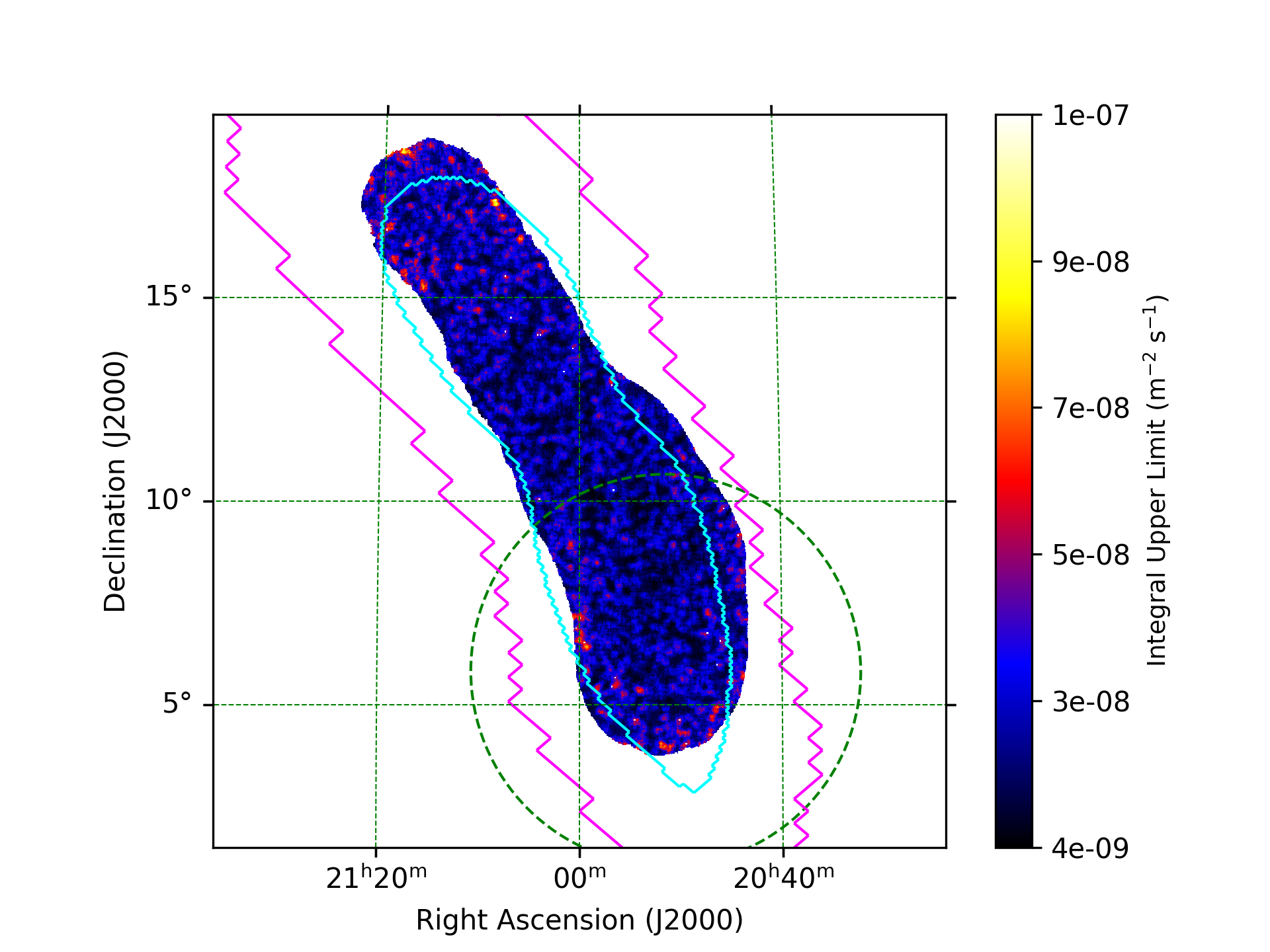}
  \end{minipage}
     \begin{minipage}[b]{0.49\textwidth}
    \includegraphics[width=\textwidth]{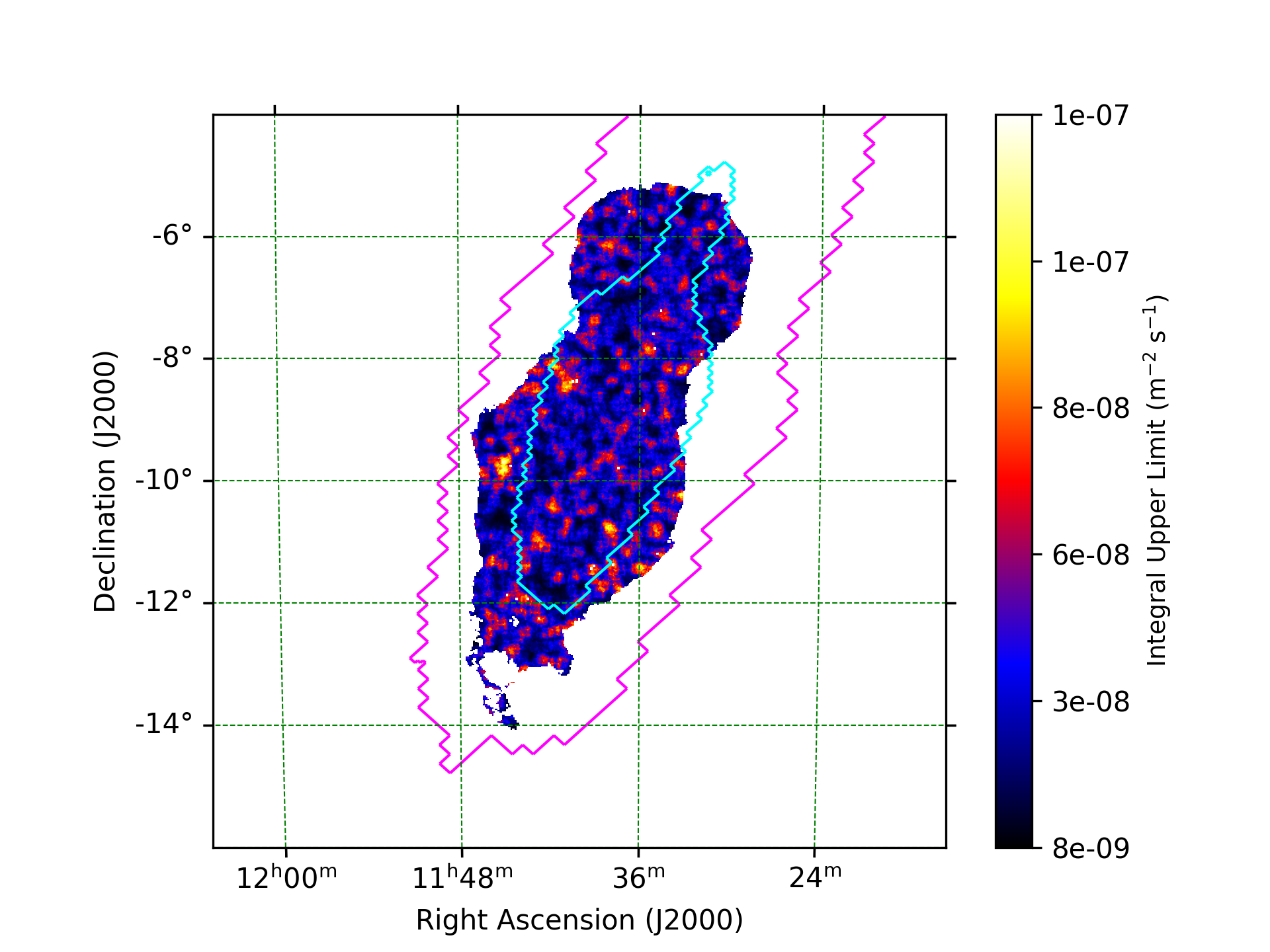}
  \end{minipage}
      \caption{Integral upper limit maps computed from the H.E.S.S. observations of GW170814 (upper left), GW190512\_180714 (upper right), GW190728\_064510 (lower left) and S200224ca (lower right) presented in Tab.~\ref{tab:HESS_OBS} and in Tab.~\ref{tab:HESS_OBS1}. We assume an internal $E^{-2}$ source spectrum taking into consideration EBL absorption effect and a specific energy range as shown in Tab.~\ref{tab:GW_REDSHIFT_INDEX_COV}. The magenta and cyan lines represent respectively the 90\% and 50\% localization contours of the newest published GW map in~\citet{Abbott_2019} for GW170814, in~\citet{abbott2020gwtc2} for GW190512\_180714 and GW190728\_064510 and in~\citet{S200224ca_update} for S200224ca. The GW maps are retrieved from~\citet{graceDB}. The green dashed circle represent the uncertainty region of the IceCube neutrino candidate~\citep{S190728q_neutrino}.}
    \label{fig:O3HESS_IntUL_Spe}
\end{figure*}

\begin{figure*}[!htb]
  \centering
  \begin{minipage}[b]{0.475\textwidth}
    \includegraphics[width=\textwidth]{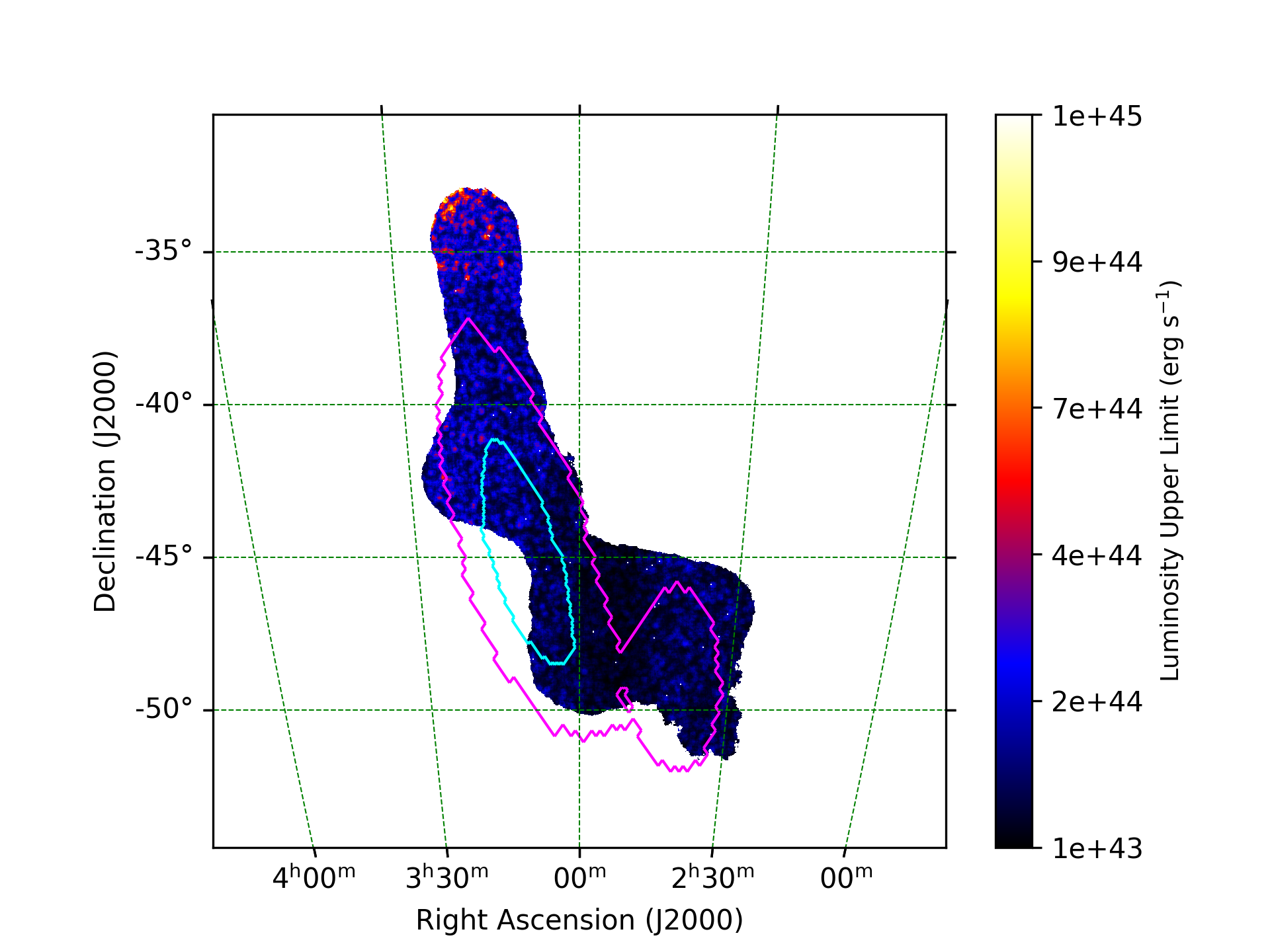}
  \end{minipage}
    \begin{minipage}[b]{0.50\textwidth}
    \includegraphics[width=\textwidth]{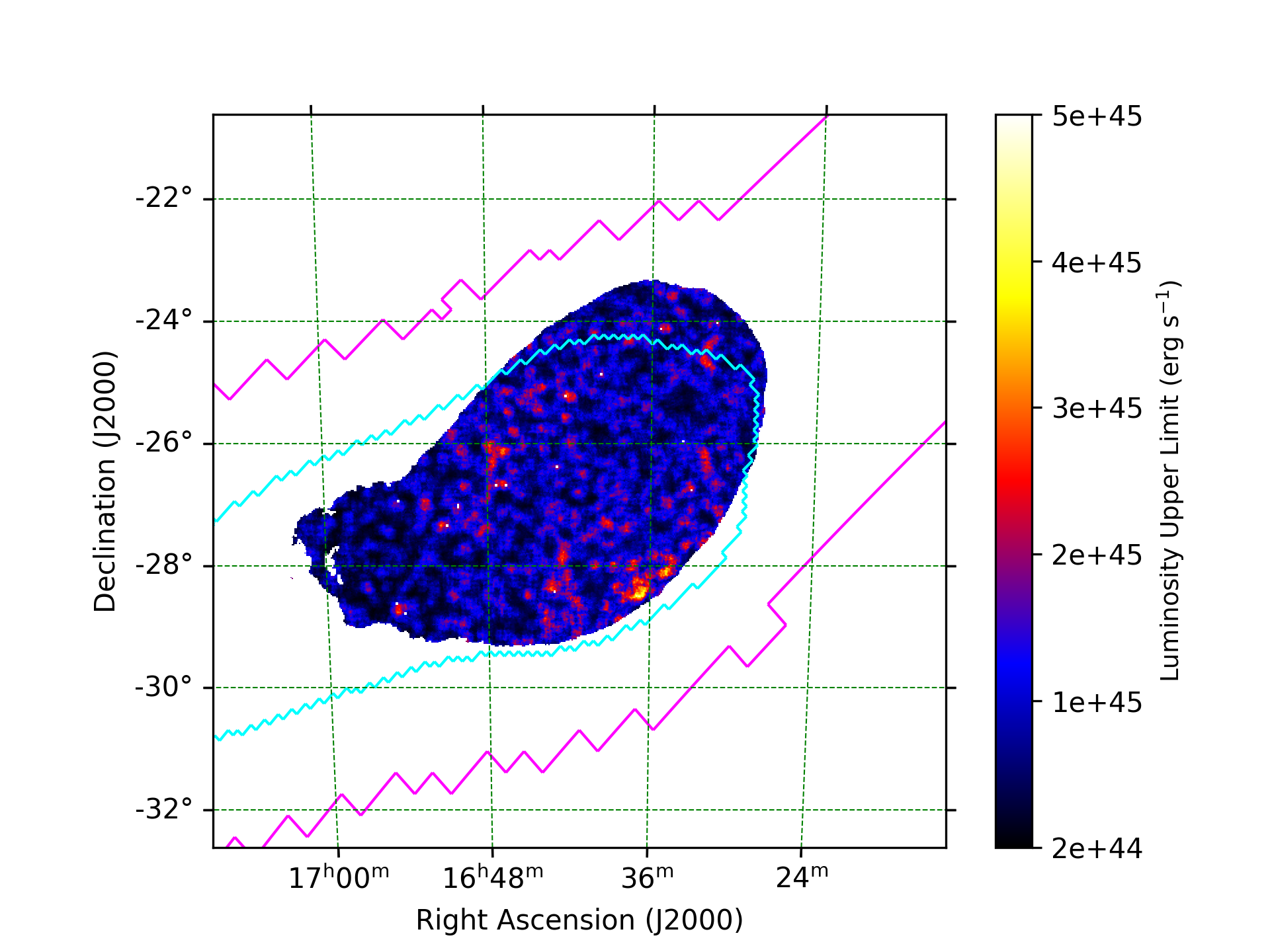}
  \end{minipage}
     \begin{minipage}[b]{0.54\textwidth}
    \includegraphics[width=\textwidth]{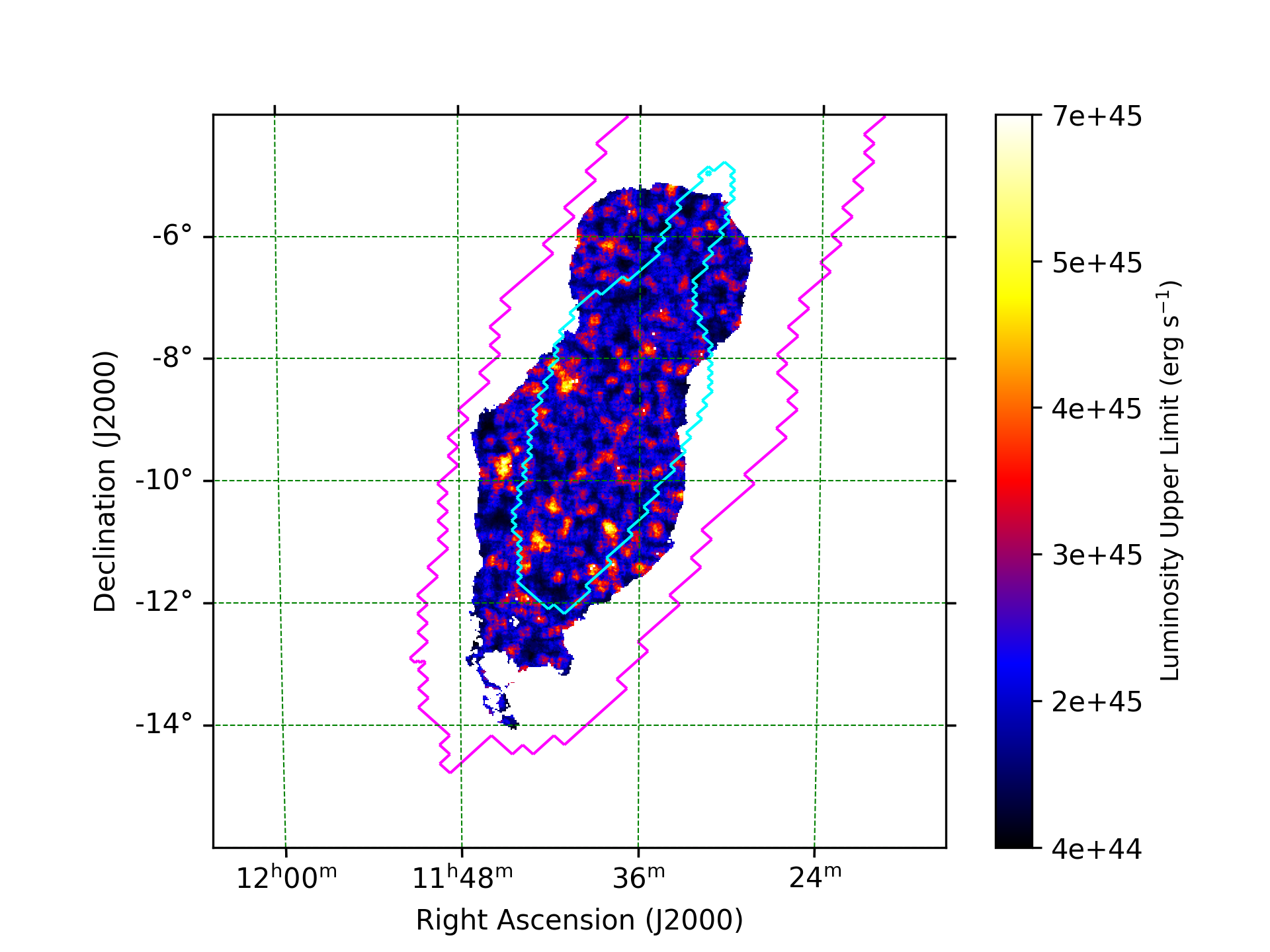}
    \label{fig:O3HESS_Lum_S200224ca}
  \end{minipage}
     \caption{Luminosity upper limit maps calculated over 1--10 TeV computed from the H.E.S.S. upper limit map for GW170814 (upper left), GW190512\_180714 (upper right) and S200224ca (lower) presented in Fig.~\ref{fig:O3HESS_IntUL_1-10} (assuming an $E^{-2}$ source spectrum) and taking into consideration the per-pixel variation of the mean distance and redshift in the newest GW maps. The magenta and cyan lines represent respectively the 90\% and 50\% localization contours of the newest published GW map in~\citet{Abbott_2019} for GW170814, in~\citet{abbott2020gwtc2} for GW190512\_180714 and in~\citet{S200224ca_update} for S200224ca. The GW maps are retrieved from~\citet{graceDB}.}
     \label{fig:O3HESS_Lum_all}
\end{figure*}

For GW170814, GW190512\_180714 and S200224ca, the pointing positions are presented in Tab.~\ref{tab:HESS_OBS}. Fig.~\ref{fig:O3HESS_IntUL_1-10} presents the integral upper limit maps for the same events in the 1-10 TeV range assuming an $E{-2}$ spectrum. Fig.~\ref{fig:O3HESS_IntUL_Spe} presents their integral upper limit maps in the specific energy range and specific spectrum indices presented in Tab.~\ref{tab:GW_REDSHIFT_INDEX_COV}. Fig.~\ref{fig:O3HESS_Lum_all} presents their luminosity integral upper limit maps. 

\begin{table*}[ht]
\centering
\small
\begin{tabular}{ccccccc}
  \hline
    Position & Start time (UTC) & RA J2000 (deg) & DEC J2000 (deg) &  Duration (min)  & Zenith angle (deg) \\  
  \hline
     1 & 2017-08-17 00:10 &  39.58 & -48.72  & 28  & 50\\
     2 & 2017-08-17 00:40 &  45.78  & -45.59  & 26 & 49\\
     3 & 2017-08-18 00:10 &  39.24  & -51.27  & 28 & 50\\
     4 & 2017-08-18 00:40 &  46.49  & -42.76  & 28 & 48\\
     5 & 2017-08-18 01:10 &  48.47  & -39.77  & 28 & 44\\
     6a & 2017-08-18 01:40 &  44.85  & -48.39 & 12 & 39\\
     6b & 2017-08-18 01:58 & 44.85  & -48.39 & 9 & 36 \\
     7 & 2017-08-18 23:56 & 38.58  & -46.20 & 28 & 50 \\
     6c & 2017-08-19 00:26 & 44.85  & -48.39 & 28 & 50 \\
     8 & 2017-08-19 00:56 & 49.24  & -37.13 & 28 & 46 \\
     9 & 2017-08-19 01:26 & 42.29  & -46.26 & 28 & 37 \\
     10 & 2017-08-19 01:56 & -34.50  & -46.26 & 28 & 34\\
     11 & 2017-08-19 02:26 & 49.86  & -42.37 & 28  & 31 \\
   \hline

  \hline
    Position & Start time (UTC) & RA J2000 (deg) & DEC J2000 (deg) &  Duration (min)  & Zenith angle (deg) \\  
      \hline
        1a & 2019-05-13 01:14 &  250.31  & -26.61  & 28 & 14 \\
     1b & 2019-05-13 01:45 &  250.31  & -26.61  & 28 & 21 \\
     2 & 2019-05-13 02:14 &  251.72  & -27.95  & 22 & 26 \\
     3a & 2019-05-13 02:38 & 248.91  & -25.28  & 20 & 34 \\
     3b & 2019-05-13 03:01 &  248.91  & -25.28  & 28 & 39 \\
     4 & 2019-05-13 03:29 &  254.53  & -27.95  & 28 & 40 \\
     \hline

  \hline
    Position & Start time (UTC) & RA J2000 (deg) & DEC J2000 (deg) &  Duration (min)  & Zenith angle (deg) \\  
      \hline
     1 & 2020-02-25 01:21 &  174.99  & -9.90  & 28 &  20 \\
     2 & 2020-02-25 01:52 &  173.58  & -6.58  & 28 &  29 \\
     3 & 2020-02-25 02:21 &  176.31  &  -13.02  & 10 &  30 \\
   \hline
    \end{tabular}
    \caption{H.E.S.S. observations of GW170814 (upper), GW190512\_180714 (middle) and S200224ca (lower) BBH merger GW events.}
    \label{tab:HESS_OBS}
\end{table*}

\newpage

\newpage

\newpage

\allauthors
\end{document}